\documentclass{aa}

\pdfoutput=1

\usepackage{graphicx}
\usepackage{txfonts}
\usepackage[colorlinks=true,urlcolor=blue,citecolor=blue]{hyperref}

\usepackage{mathtools}
\DeclarePairedDelimiter\ceil{\lceil}{\rceil}
\DeclarePairedDelimiter\floor{\lfloor}{\rfloor}

\usepackage{etoolbox}
\makeatletter
\patchcmd\@combinedblfloats{\box\@outputbox}{\unvbox\@outputbox}{}{%
   \errmessage{\noexpand\@combinedblfloats could not be patched}%
}%
 \makeatother

\begin{document}

\title{Data calibration for the MASCARA and bRing instruments}

\author{
G.J.J. Talens \inst{1}
\and E.R. Deul \inst{1}
\and R. Stuik \inst{1}
\and O. Burggraaff \inst{1,4}
\and A.-L. Lesage \inst{1}
\and J.F.P. Spronck \inst{1}
\and S. N. Mellon \inst{2}
\and J.I. Bailey, III \inst{1}
\and E. E. Mamajek \inst{3,2}
\and M.A. Kenworthy \inst{1}
\and I.A.G. Snellen \inst{1}
}

\institute{Leiden Observatory, Leiden University, PO Box 9513, 2300 RA, Leiden, The Netherlands\\
\email{talens@strw.leidenuniv.nl}
\and Department of Physics \& Astronomy, University of Rochester, 500 Wilson Blvd., Rochester, NY 14627-0171, USA
\and Jet Propulsion Laboratory, California Institute of Technology, M/S 321-100, 4800 Oak Grove Drive, Pasadena, CA 91109, USA
\and Institute of Environmental Sciences (CML), Leiden University, PO Box 9518, 2300 RA, Leiden, The Netherlands}

\abstract
{}
{MASCARA and bRing are photometric surveys designed to detect variability caused by exoplanets in stars with $m_V < 8.4$. Such variability signals are typically small and require an accurate calibration algorithm, tailored to the survey, in order to be detected. This paper presents the methods developed to calibrate the raw photometry of the MASCARA and bRing stations and characterizes the performance of the methods and instruments.}
{For the primary calibration a modified version of the coarse decorrelation algorithm is used, which corrects for the extinction due to the earth's atmosphere, the camera transmission, and intrapixel variations. Residual trends are removed from the light curves of individual stars using empirical secondary calibration methods. In order to optimize these methods, as well as characterize the performance of the instruments, transit signals were injected in the data.}
{After optimal calibration an RMS scatter of 10 mmag at $m_V \sim 7.5$ is achieved in the light curves. By injecting transit signals with periods between one and five days in the MASCARA data obtained by the La Palma station over the course of one year, we demonstrate that MASCARA La Palma is able to recover 84.0, 60.5 and 20.7\% of signals with depths of 2, 1 and 0.5\% respectively, with a strong dependency on the observed declination, recovering 65.4\% of all transit signals at $\delta > 0 \degr$ versus 35.8\% at $\delta < 0 \degr$. Using the full three years of data obtained by MASCARA La Palma to date, similar recovery rates are extended to periods up to ten days. We derive a preliminary occurrence rate for hot Jupiters around A-stars of ${>} 0.4 \%$, knowing that many hot Jupiters are still overlooked. In the era of TESS, MASCARA and bRing will provide an interesting synergy for finding long-period (${>} 13.5 \rm{~days}$) transiting gas-giant planets around the brightest stars.}
{}

\maketitle

\section{Introduction}
\label{sec:introduction}

Exoplanet transit survey instruments, as with all astronomical instrumentation, have to take care of systematic effects in the data. Some of these systematics, such as airmass effects, clouds and point spread function (PSF) modulations are present in many of such surveys and general methods have been developed for their removal, for example \textsc{sysrem} \citep{Tamuz2005} and the trend filtering algorithm \citep[TFA;][]{Kovacs2005}. Both of these methods use an ensemble of observed stars to remove systematics that are common in many light curves. Other systematics are unique to specific surveys, such as shutter shadows \citep[SuperWASP;][]{CollierCameron2006}, scattered light from earth \citep[CoRoT;][]{Grziwa2012} and cosmic ray impact sensitivity variations \citep[\emph{Kepler};][]{Stumpe2012}. The presence of these survey specific systematics means that, rather than using a general pipeline, a unique calibration pipeline is written for most surveys. Nevertheless, many of these pipelines are built around the same principle, using ensembles of stars to remove common systematics. The \emph{Kepler} pre-search data conditioning \citep[PDC;][]{Stumpe2012,Smith2012} is a well-known example of such a pipeline. The data obtained by the MASCARA and bRing instruments also contains instrument specific systematics, caused by the stars moving across the detectors during the night. In this paper we describe the systematics present in the data obtained by the MASCARA and bRing surveys, and the calibration pipeline used to remove them.

The Multi-site All-Sky CAmeRA \citep[MASCARA;][]{Talens2017a} consists of two observing stations, one located at the Observatorio del Roque de los Muchachos in the northern hemisphere, and one located at La Silla observatory in the southern hemisphere. Each station employs five cameras, pointed at the cardinal directions and zenith, to observe all bright stars ($m_V < 8.4$) down to airmass three in order to detect transiting exoplanets. These exoplanets make ideal targets for atmospheric characterization at high spectral resolution. To date two new exoplanets have been found with MASCARA \citep{Talens2017b,Talens2018a,Lund2017}, with several more candidates still awaiting confirmation.

\begin{table*}
\centering
\caption{Summary of observing stations.}
\begin{tabular}{l l l l l}
 Instrument & Observatory & Site ID & Cameras & Camera IDs\tablefootmark{a} \\
 \hline
 \hline
 MASCARA & Roque de los Muchachos & LP (La Palma) & 5 & N,E,S,W,C \\
 MASCARA & La Silla & LS (La Silla) & 5 & N,E,S,W,C \\
 bRing & Sutherland & SA (South Africa) & 2 & E,W \\
 bRing & Siding Spring & AU (Australia) & 2 & E,W
\end{tabular}
\tablefoot{
\tablefoottext{a}{The cameras pointing towards zenith are identified with a `C' for `central'.}
}
\label{tab:stations}
\end{table*}

The $\beta$ Pictoris b Ring project \citep[bRing;][]{Stuik2017} consists of two observing stations in the southern hemisphere, located at the Sutherland observing station and Siding Spring observatory. Each station employs two cameras, pointed to the east and west, to observe all bright stars ($m_V < 8.4$) with $\delta \lesssim -30 \degr$. The main goal of bRing is to monitor the star $\beta$ Pictoris for the predicted 2017/2018 Hill Sphere transit of the directly imaged exoplanet $\beta$ Pic\,b in an attempt to detect circumplanetary material, such as a ring system similar to the one found around the exoplanet J 1407 b \citep{Mamajek2012}. Furthermore, bRing searches for evidence of stellar variability in the bright stars it monitors (Mellon et al., in prep).

Together MASCARA and bRing monitor all bright stars in the sky with a total of 14 cameras and provide near continuous coverage for stars with $\delta \lesssim -30 \degr$; both surveys are sensitive to evidence of stellar variability from sources such as exoplanets \citep{Talens2017b,Talens2018a} and pulsations (Mellon et al., in prep). For convenience, each of the 14 cameras is referred to by a three letter identifier, with the first two letters identifying the observing site and the last letter the camera by its pointing direction, for example LSE for the east camera of the station at La Silla observatory, see Table \ref{tab:stations} for the other identifiers.

The stations take synchronized observations of the sky at $6.4$ sidereal second intervals for a total of $13500$ observing slots per sidereal day. Each MASCARA exposure lasts the full $6.4$ sidereal seconds, while bRing alternates between taking $6.4$ and $2.55$ sidereal second exposures, to accommodate the bright ($m_V = 3.86$) $\beta$ Pic system. Each exposure is uniquely identified by its \textsc{lstseq}\footnote{The \textsc{lstseq} is defined from the local sidereal time at La Palma starting from $\textsc{lst} = 0 \rm{~h}$ at UTC 2012-12-31 18:29:10.68}, an index indicating the $6.4$ second exposure slot during which it was taken, and exposures taken at the same local sidereal time are identified using the \textsc{lstidx}, defined as $\textsc{lstidx} = (\textsc{lstseq} \mod 13500)$.

MASCARA and bRing exposures are processed by an on-site reduction pipeline which performs astrometry, creates stacked images and performs photometry. The MASCARA observations were originally designed for difference imaging photometry \citep{Talens2017a}; however, difference imaging algorithms were found to be unable to keep pace with the data acquisition rate and aperture photometry is used instead. The aperture photometry is performed on stars with $m_V < 8.4$ in the raw images, producing light curves at $6.4 \rm{~s}$ cadence (fast light curves), and on stars with $m_V < 10$ in the stacked images, producing light curves at $320 \rm{~s}$ cadence (slow light curves). The reduced data are then transferred to Leiden for calibration. 

This paper details the calibration of the fast light curves, describing the primary calibration of the data in Sect. \ref{sec:primary_cal}, and a secondary calibration, tailored to find short duration periodic features such as transiting exoplanets, in Sect. \ref{sec:secondary_cal}. The transit detection algorithm is described in Sect. \ref{sec:detection} and signal recovery test are presented in Sect. \ref{sec:recovery}. In Sect. \ref{sec:occurrence} we present an estimate of the occurrence rate of hot Jupiters around A-stars. Finally, in Sect. \ref{sec:conclusions} we summarize the results and present our conclusions. 

\section{Primary calibration}
\label{sec:primary_cal}

\begin{figure*}
  \centering
  \includegraphics[width=8.5cm]{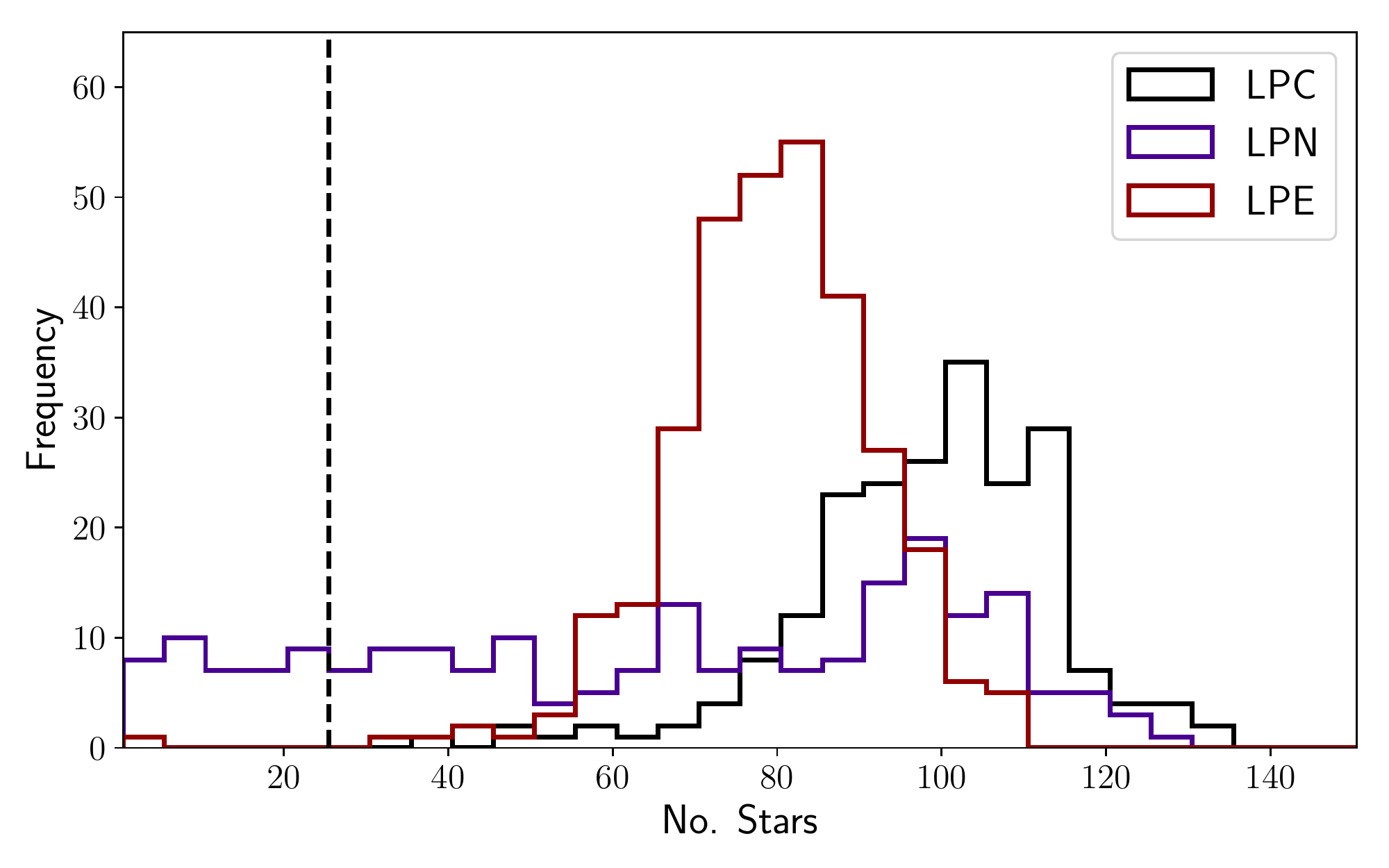}
  \includegraphics[width=8.5cm]{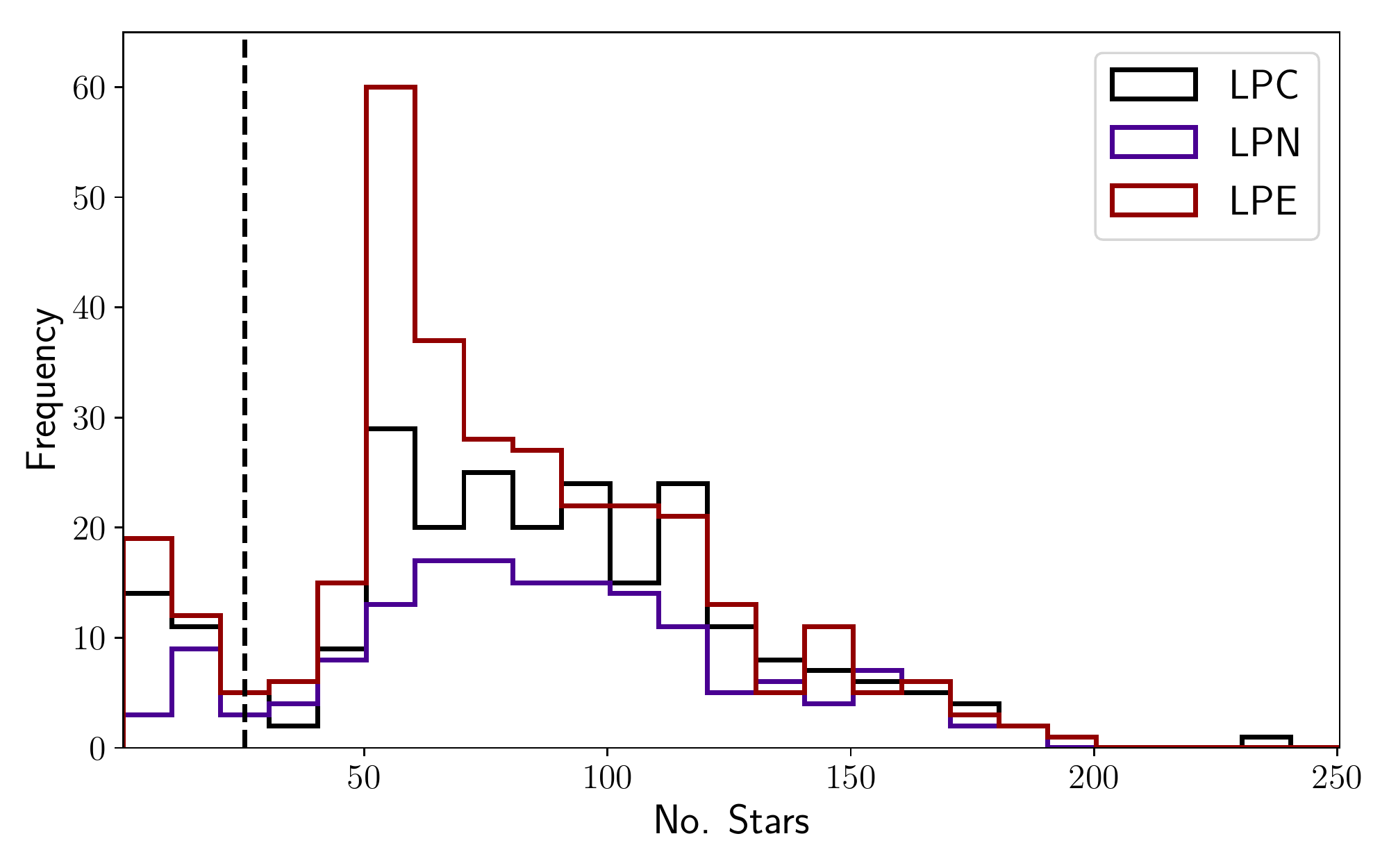}
  \caption{Histograms illustrating how the number of stars observed sets the resolution of the Polar and HEALPix grids for the LPC, LPN and LPE cameras during the 2016-10-A baseline, the black dotted line indicates $25$ stars per resolution element. \textit{Left:} frequency of rings in the polar grid containing a certain number of stars. \textit{Right:} frequency of patches in the HEALPix grid containing a certain number of stars.}
  \label{fig:hist}
\end{figure*}

Three main systematics are common to all MASCARA and bRing photometry \citep[see also Fig. 10 of][]{Talens2017a}. First, extinction due to the earth's atmosphere, such as clouds, Calima and airmass effects. Second, the changing transmission of the lenses and changes in the fraction of light in the photometric aperture due to changes in the PSF shape with position on the CCD. Third, the intrapixel variations, a sinusoidal modulation due to the interline design of the CCDs. 

We correct these effects for each of the cameras individually using baselines of ${\sim} 15 \rm{~days}$, running from the 1st to the 15th and from the 16th to the end of each month, denoted `A' and `B' respectively. These baselines are long enough to disentangle the atmospheric and camera dependent effects even when several nights are dominated by the atmosphere, while simultaneously short enough to keep the memory required for calibration reasonable. For each baseline data flagged by the on-site reduction\footnote{Some flags are not implemented in the on-site reduction of MASCARA La Palma and are performed on the fly instead.} is removed and the raw flux measurements and uncertainties are converted to magnitudes, using

\begin{align}
m_{it} = -2.5\log_{10}\Bigg(\frac{F_{it}}{t_{\rm{exp},t}}\Bigg) + 25.
\end{align}

In this equation $F_{it}$ is the raw flux in analog-to-digital units (ADUs), $m_{it}$ the magnitude and $t_{\rm{exp},t}$ the corresponding exposure time. The offset of $25$ magnitudes is arbitrary and for the MASCARA station on La Palma the division by the exposure time is not performed, as this only became necessary after the conception of bRing with its alternating exposure times. The index $i$ identifies the star and the index $t$ is short for the \textsc{lstseq}. The combination of $i$ and $t$ uniquely identifies each data point for a particular camera. Subsequently we correct for the main systematics using a variation on the coarse decorrelation algorithm, proposed by \citet{CollierCameron2006}, modified to include the effects of the transmission and intrapixel variations, which leads to the log-likelihood equation:

\begin{align}
\ln L = -\frac{1}{2} \sum_{i,t} \bigg[&\ln(2\pi(\sigma_{it}^2 + \sigma_i^2 + \sigma_{qt}^2)) + \nonumber \\
&\frac{(m_{it}  - m_i - c_{qt} - T_{nk} - f(x_{it}, y_{it}))^2}{\sigma_{it}^2 + \sigma_i^2 + \sigma_{qt}^2}\bigg].\label{eq:loglike}
\end{align}

In this equation $m_{it}$ and $\sigma_{it}$ are the raw magnitudes and uncertainties. We describe these data as a sum of $m_i$, the mean magnitude of the star, $c_{qt}$, the atmospheric correction, $T_{nk}$, the transmission correction and $f(x_{it}, y_{it})$, a function describing the intrapixel variations. For the function $f(x_{it}, y_{it})$ we adopt the form,

\begin{align}
f(x_{it},y_{it}) =& a_{nl}\sin(2\pi x_{it}) + b_{nl}\cos(2\pi x_{it}) + \nonumber \\
& c_{nl}\sin(2\pi y_{it}) + d_{nl}\cos(2\pi y_{it}),\label{eq:intrapixel}
\end{align}

describing the intrapixel variations as sinusoids in the position on the CCD $x_{it}, y_{it}$, with amplitudes $a_{nl}, b_{nl}, c_{nl}, d_{nl}$. Following \citet{CollierCameron2006} we also solve for additional uncertainties $\sigma_i$, which reduces the influence of variable stars and residual trends (see Sect. \ref{sec:secondary_cal}), and $\sigma_{qt}$, which reduces the effects of cloudy data on the overall solution. The inclusion of these additional uncertainties eliminates the need for additional data selection or sigma clipping. 

The indices $q,n,k$ and $l$ indicate the spatial dependence of these corrections, which are solved onto discretized grids. These indices, obtained from quantities uniquely identified by $i$ and $t$, will be described in detail in the next sections. Equation \ref{eq:loglike} is iteratively solved for $T_{nk}$ and $a_{nl},b_{nl},c_{nl},b_{nl}$, which we will refer to as the spatial component, and $m_i, \sigma_i$ and $c_{qt}, \sigma_{qt}$, the temporal component. 

\subsection{Spatial}

During most nights the effects of the lens transmission and intrapixel variations are the dominant systematics in the MASCARA and bRing light curves, as such we start the iterative procedure by solving for the spatial corrections. For convenience we define $\bar{m}_{it} = m_{it} - m_i - c_{qt}$ and $\bar{\sigma}^2_{it} = \sigma^2_{it} + \sigma^2_i + \sigma^2_{qt}$ so that we may rewrite eq. \ref{eq:loglike} as

\begin{equation}
\ln L = -\frac{1}{2}\sum_{i,t} \bigg[\ln(2\pi\bar{\sigma}_{it}^2) + \frac{(\bar{m}_{it} - T_{nk} - f(x_{it}, y_{it}))^2}{\bar{\sigma}_{it}^2}\bigg],\label{eq:loglike_spatial}
\end{equation}

and solve for the spatial corrections. For the spatial dependence of transmission $T_{nk}$, we use a polar grid in hour angle (\textsc{ha}) and declination ($\delta$) where 

\begin{align}
k = \ceil*{\frac{\textsc{ha}_{it}}{6.4 \rm{~s}}} \text{ and } n = \ceil*{\frac{\delta_i + 90\degr}{0.25\degr}}.
\end{align}

The resolution along the hour angle axis emerges naturally from the $6.4 \rm{~s}$ observational cadence and fixed sidereal observations times employed by MASCARA and bRing and is therefore set to $6.4 \rm{~s}$. The resolution along the declination axis is motivated by the requirement that there is a sufficient number of stars in each declination ring $n$ while not making the rings too wide to properly sample the transmission. After testing several resolutions we decided on rings with a width of $0.25\degr$ for a total of $720$ rings across the sky. Figure \ref{fig:hist} shows a histogram of the number of stars in each ring for the zenith, north and east cameras of MASCARA La Palma (LPC, LPN, LPE) during the first half of October 2016 (baseline 2016-10-A), similar distributions are obtained for other cameras and baselines. For the LPC and LPE cameras the majority of the rings contain ${>} 50$ stars, resulting in reliable correction terms. For the LPN camera the distribution has a long tail towards zero, a consequence of the decreasing area of each ring as we move towards the pole. 

For the intrapixel variations, described by $a_{nl},b_{nl},c_{nl},d_{nl}$ we also use a polar grid in hour angle and declination, employing the same index $n$ along the declination axis while defining
 
\begin{align}
l = \ceil*{\frac{\textsc{ha}_{it}}{320 \rm{~s}}},
\end{align}

along the hour angle axis. By setting a resolution of $320 \rm{~s}$, a factor 50 lower than the resolution used for the transmission map, we ensure the cells $nl$ contain a sufficient range of $x$, $y$ to properly sample the sinusoids.

\begin{figure*}
  \centering
  \includegraphics[width=17cm]{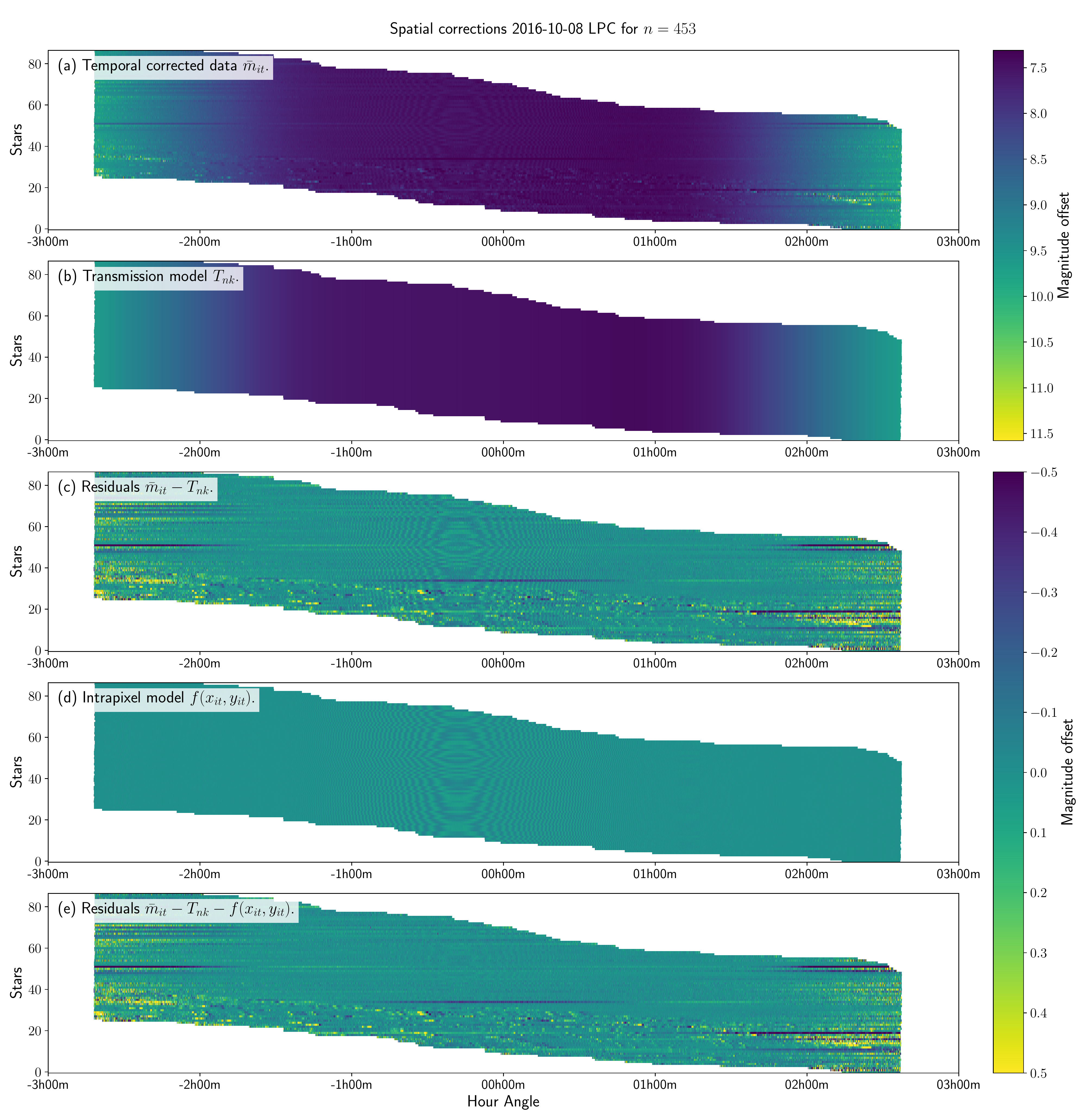}
  \caption{Data, model and residuals for ring $n=543$ of the LPC camera during the 2016-10-A baseline. For clarity only the night of 2016-10-08 is shown and the median of the residual rows has been subtracted from all panels, removing small normalization differences due to keeping the $m_i$ fixed. \textit{(a)} Photometric time-series for the stars in ring $n=453$ during the night of 2016-10-08 as a function of the hour angle after subtracting the mean magnitude $m_{i}$ and cloud corrections $c_{qt}$. \textit{(b)} The transmission corrections $T_{nk}$ obtained from solving Eq. \ref{eq:loglike} \textit{(c)} The residuals after subtracting the transmission corrections. \textit{(d)} The intrapixel corrections $f(x_{it}, y_{it})$ obtained from solving Eq. \ref{eq:loglike}. \textit{(e)} The residuals after subtracting the intrapixel corrections.}
  \label{fig:spatial}
\end{figure*}

\subsection{Temporal}

After obtaining corrections for the lens transmission and intrapixel variations we solve for the atmospheric extinction corrections. Let us define $\tilde{m}_{it} = m_{it} - T_{nk} - f(x_{it},y_{it})$ so that we may rewrite eq. \ref{eq:loglike} as 

\begin{equation}
\ln L = -\frac{1}{2}\sum_{i,t} \bigg[\ln(2\pi(\sigma_{it}^2 + \sigma_i^2 + \sigma_{qt}^2)) + \frac{(\tilde{m}_{it} - m_i - c_{qt})^2}{\sigma_{it}^2 + \sigma_i^2 + \sigma_{qt}^2}\bigg].\label{eq:loglike_temporal}
\end{equation}

In this form the equation most closely resembles that of the coarse decorrelation presented by \citet{CollierCameron2006}. However, the large field of view (FoV) of the individual cameras ($53\degr\times74\degr$) means the implicit assumption made by \citet{CollierCameron2006} that any atmospheric corrections can be averaged over the FoV breaks down. In order to take this into account we subdivide the sky into patches of right ascension and declination identified by the index $q$. For this division of the sky we use the Hierarchical Equal Area isoLatitude Pixelization \citep[HEALPix;][]{Gorski2005}. As with the declination rings our choice for the resolution of the HEALPix grid, set by the $N_{\rm{side}}$ parameter, is motivated by the requirement that there is a sufficient number of stars in each patch $q$ while also optimally sampling the position dependence of the atmospheric extinction. Based on this we use a HEALPix grid with $N_{\rm{side}} = 8$, resulting in $768$ patches across the sky, each with an area of $53.7 \rm{~deg}^2$. Figure \ref{fig:hist} shows a histogram of the number of stars in each patch for the LPC, LPN and LPE cameras during the 2016-10-A baseline. As for the rings most patches contain ${>} 50$ stars, resulting in reliable correction terms. In addition, the design of the HEALPix grid avoids the polar singularity of the polar grid. However, due to the projection of the HEALPix grid onto the CCDs some patches are never fully observed, causing a secondary peak towards zero for all cameras.

\subsection{Solution and caveats}

\begin{figure*}
  \centering
  \includegraphics[width=17cm]{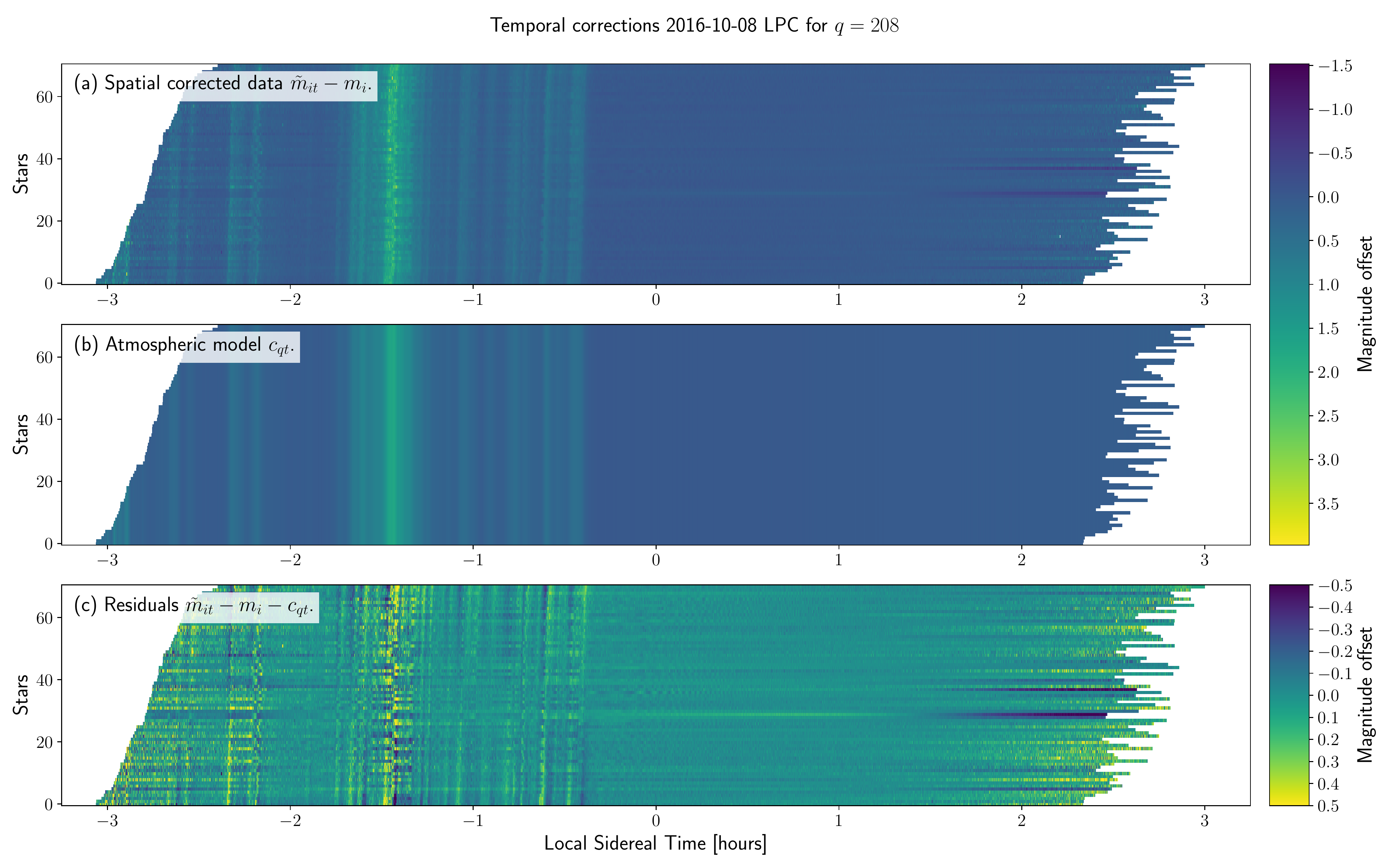}
  \caption{Data, model and residuals for patch $q=208$ of the LPC camera during the 2016-10-A baseline. For clarity only the night of 2016-10-08 is shown and the median of the residual rows has been subtracted from all panels, removing small normalization differences due to keeping the $m_i$ fixed. \textit{(a)} Photometric time-series for the stars in patch $q=208$ during the night of 2016-10-08 as a function of local sidereal time after subtracting the transmission and intrapixel corrections $T_{nk}$ and $f(x_{it}, y_{it})$. For clarity the mean magnitude $m_{i}$ has also been removed. \textit{(b)} The atmospheric corrections $c_{qt}$ obtained from solving Eq. \ref{eq:loglike}. \textit{(c)} The residuals after subtracting the atmospheric corrections.}
  \label{fig:temporal}
\end{figure*}

We solve Eq. \ref{eq:loglike_spatial} and Eq. \ref{eq:loglike_temporal} iteratively, initially setting $c_{qt} = 0$, $\sigma_i = 0$ and $\sigma_{qt} = 0$. Furthermore, we fix the mean magnitudes to the known visual magnitudes of the stars ($m_{i} = m_{V,i}$) in order to keep the reference level the same over time-scales longer than the individual baselines. Because the indices $q$ and $n$ are solely dependent on quantities related to the stars $i$ we can further solve individual rings $n$ and individual patches $q$ independently when solving Eq. \ref{eq:loglike_spatial} and Eq. \ref{eq:loglike_temporal}, respectively. However, in the full solution all corrections depend on one another to some degree as stars in the same ring $n$ will contribute to different patches $q$ and vice versa. Further details on how Eqs. \ref{eq:loglike_spatial} and \ref{eq:loglike_temporal} are solved can be found in appendix \ref{app:solution}.

Figure \ref{fig:spatial} illustrates the spatial corrections, in the shape of the solution of Eq. \ref{eq:loglike_spatial}, for the LPC camera during the 2016-10-A baseline for a particular ring $n$. Panel (a) shows the photometry of all stars in the ring as function of hour angle, after subtracting the temporal corrections. The stars are ordered according to the starting time of their observations. At the beginning of the night there were some thin clouds and though the atmospheric correction has already been subtracted there is more noise on the stars observed early in the night. Panel (b) shows the transmission corrections $T_{nk}$. Panel (c) shows the residuals after subtracting the transmission corrections from the data. The noise due to clouds is now more evident, and the fringe-like pattern introduced by the intrapixel variations has become visible, peaking between \textsc{ha} $-00h30m$ and $00h00m$. Panel (d) shows the intrapixel corrections $f(x_{it}, y_{it})$. We note here that the free parameters $a_{nl}$, $b_{nl}$, $c_{nl}$ and $d_{nl}$ depend only on the hour angle for fixed $n$, all other structure comes from the positions of the stars on the CCD $(x_{it},y_{it})$ and is not the result of optimizing free parameters. Panel (e) shows the residuals after subtracting both the transmission and intrapixel corrections from the data.

\begin{figure*}
  \centering
  \includegraphics[width=17cm]{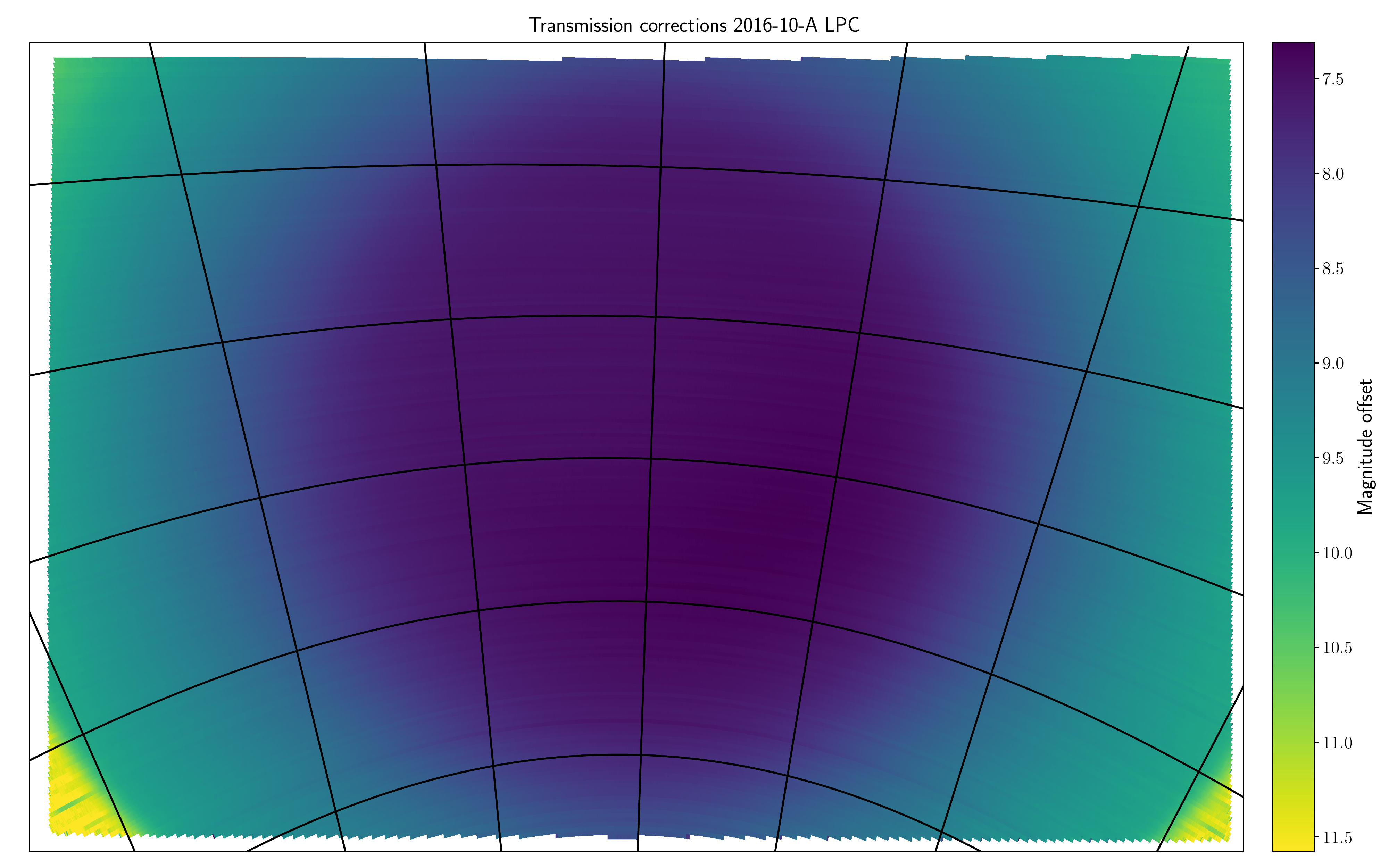}
  \caption{The transmission map obtained from solving Eq. \ref{eq:loglike} with all data obtained with the LPC camera during the 2016-10-A baseline. The maps are shown projected onto the CCD and a grid in hour angle and declination is overlaid on top.}
  \label{fig:transmission}
\end{figure*}

\begin{figure*}
  \centering
  \includegraphics[width=17cm]{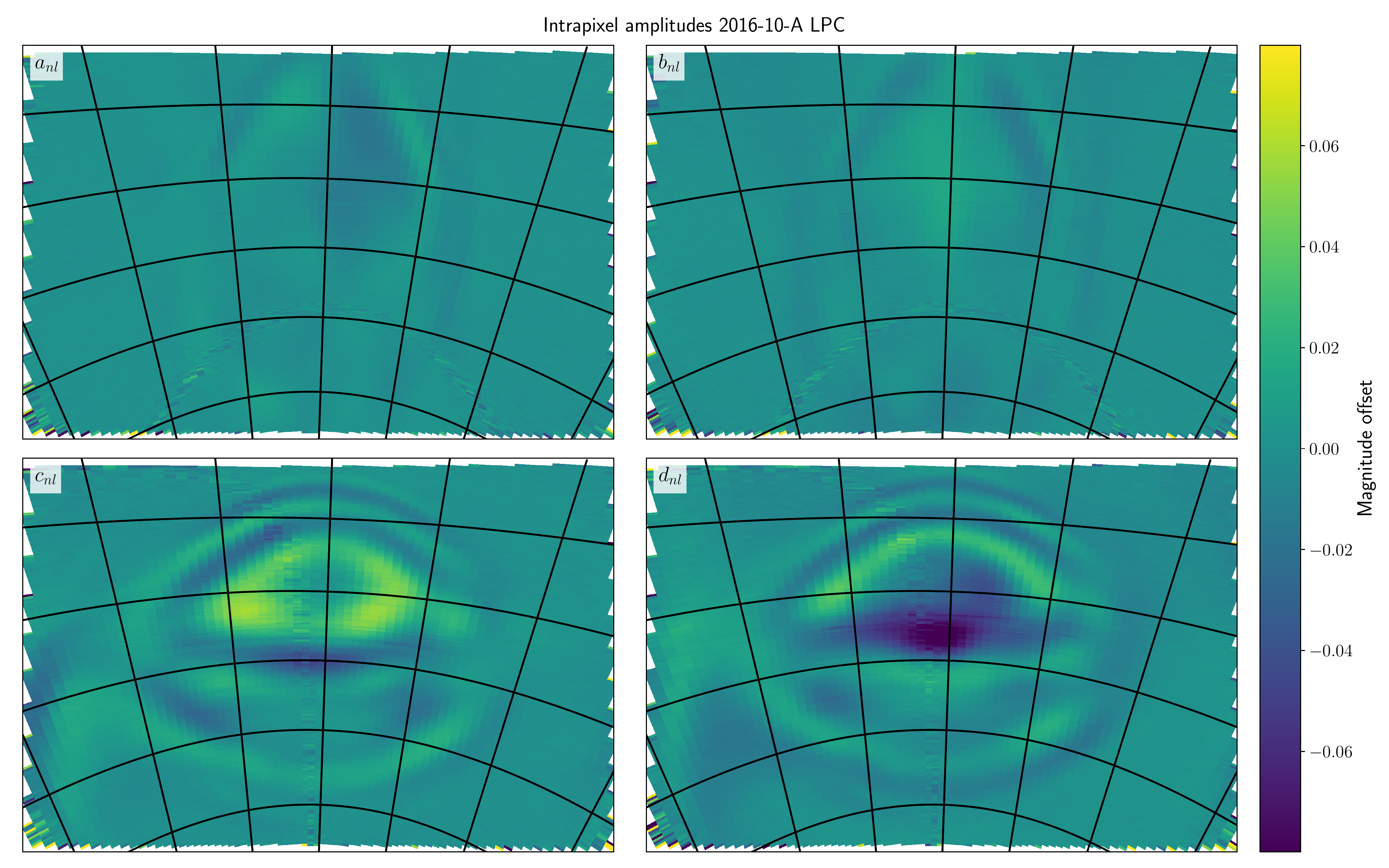}
  \caption{The intrapixel amplitude maps obtained from solving Eq. \ref{eq:loglike} with all data obtained with the LPC camera during the 2016-10-A baseline. The maps are shown projected onto the CCD and a grid in hour angle and declination is overlaid on top. The top panels show the amplitudes $a_{nl}$ and $b_{nl}$ of the $\sin(2\pi x_{it})$ and $\cos(2\pi x_{it})$ terms, the bottom panels show the amplitudes $c_{nl}$ and $d_{nl}$ of the $\sin(2\pi y_{it})$ and $\cos(2\pi y_{it})$ terms.} 
  \label{fig:intrapixel}
\end{figure*}

Figure \ref{fig:temporal} illustrates the temporal corrections, in the shape of the solution of Eq. \ref{eq:loglike_temporal}, for the LPC camera during the 2016-10-A baseline for a particular patch $q$. Panel (a) shows the photometry of all stars in the sky-patch as a function of sidereal time, after subtracting the spatial corrections and the mean magnitudes. Several nearly vertical features are present at times when the stars were obscured by thin clouds. Panel (b) shows the atmospheric extinction corrections $c_{qt}$. Panel (c) shows the residuals after subtracting the atmospheric extinction correction from the data. The residuals are slightly skewed, caused by motion of the clouds across the sky-patch and extinction gradients in the clouds covering the sky-patch. In both Fig. \ref{fig:spatial} and Fig. \ref{fig:temporal} individual stars show residuals trends, the possible origins of these trends are discussed in Sect. \ref{sec:secondary_cal}. 

\begin{figure*}
  \centering
  \includegraphics[width=17cm]{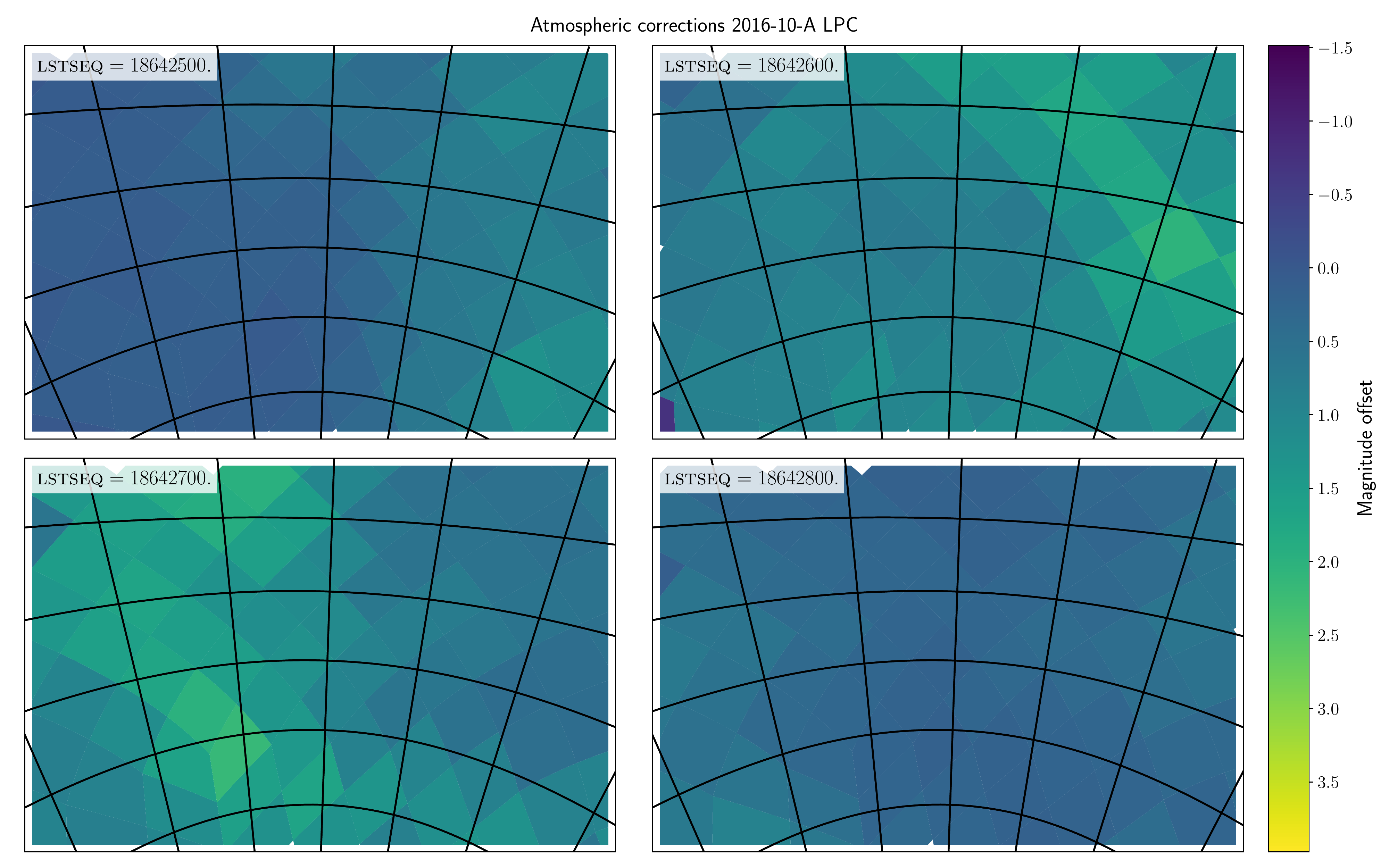}
  \caption{The atmospheric corrections obtained from solving Eq. \ref{eq:loglike} with all data obtained with the LPC camera during the 2016-10-A baseline. Each panel shows the solution at a different \textsc{lstseq} during the night of 2016-10-08 with an interval of 100 exposures or $640$ sidereal seconds. The maps are shown projected onto the CCD and a grid in hour angle and declination is overlaid on top. A movie of the atmospheric corrections during the night of 2016-10-08 is available online (see footnote \ref{note:movie}). }
  \label{fig:clouds}
\end{figure*}

Figure \ref{fig:transmission} shows the full transmission map for the LPC camera during the 2016-10-A baseline. The on-sky projection of this camera is representative of the LPC, LPS, LSC and LSN cameras of the MASCARA stations. For these cameras stars move approximately horizontally along the CCD. Small variations in the normalization of the different declination rings $n$ are present because the photometry only truly samples the transmission along the direction of motion of the stars and the normalization between declination rings is imposed by setting the magnitudes of the stars. 

Figure \ref{fig:intrapixel} shows the full intrapixel amplitude maps for the LPC camera during the 2016-10-A baseline. For this particular camera the intrapixel variations depend more strongly on the $y$ position, but in general they depend more strongly on either $x$ or $y$. In the top row ($x$ dependence) an arc is visible at the bottom of the CCD where the quality of the solution decreases, similarly in the bottom row ($y$ dependence) a curved vertical area near the centre of the CCD is degraded. In these regions of the CCD the change in the $x$ or $y$ position of a star between exposures is either close to $1$ pixel or less than $1/50$th of a pixel, resulting in a poor sampling of the sinusoidal modulations in the cell $nl$. Such structures are present in the intrapixel maps of all MASCARA and bRing cameras, with their shapes determined by the on-sky orientation of the camera. Examples of the transmission and intrapixel maps for other on-sky orientations are shown in appendix \ref{app:extra_maps}.

\begin{figure*}
  \centering
  \includegraphics[width=17cm]{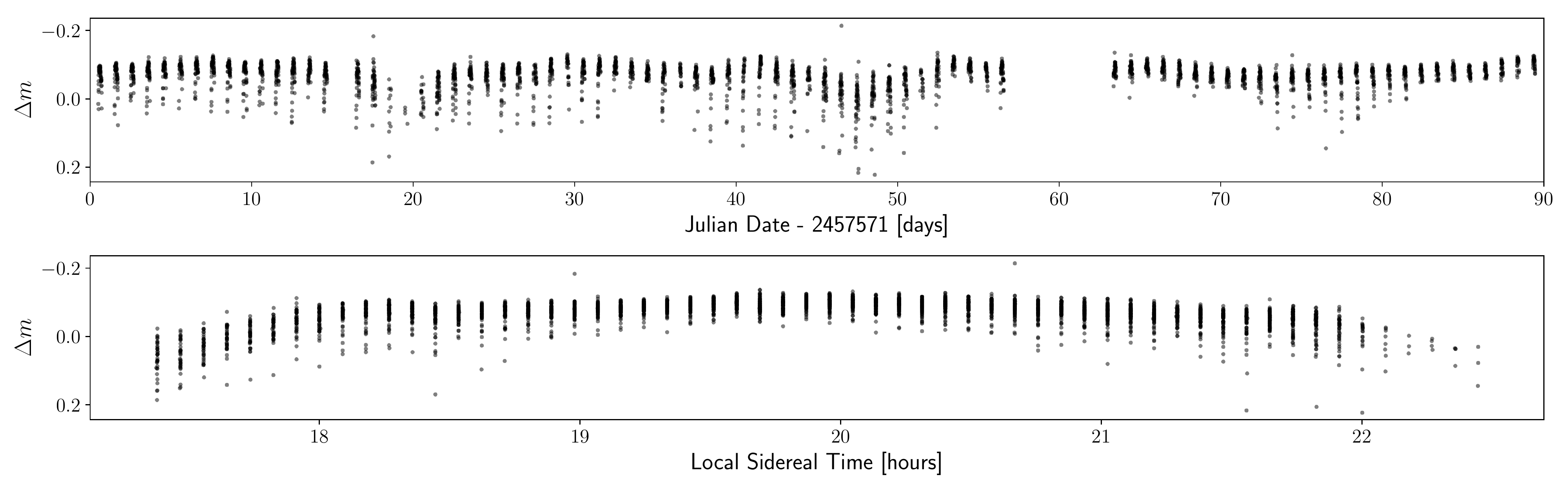}
  \caption{Light curve of HD\,189733 as observed by the LPC camera during 2016Q3. The top panel shows the light curve as a function of the Julian date, highlighting the long-term variations, in particular the effect of the moon is visible at ${\sim} 20$, ${\sim} 48$ and ${\sim} 76 \rm{~days}$. The bottom panel shows the light curve as a function of the local sidereal time, highlighting the PSF effect.}
  \label{fig:trends}
\end{figure*}

Figure \ref{fig:clouds} shows the atmospheric corrections for LPC camera during the 2016-10-A baseline at four different times, and a movie of the atmospheric extinction map during the night of 2016-10-08 is available online\footnote{The movie associated to Fig. \ref{fig:clouds} is available at \url{http://www.aanda.org} \label{note:movie}}. The stars move across the LPC camera from left-to-right, as seen by the change in the location of the sky-patches, while some thin clouds moved across the FoV from right-to-left.

After computing the correction terms, calibrated light curves are obtained by subtracting the correction terms from the raw magnitudes. Next, the photometry is flagged and removed if ${<} 25$ points were used in computing $c_{qt}$, $T_{nk}$ or $a_{nl},b_{nl},c_{nl},d_{nl}$ in order to remove data where the corrections are poorly constrained. This removes rings and patches to the left of the black dashed line in Fig. \ref{fig:hist}. In addition we flag and remove photometry if $\sigma_{qt} > 0.05$, to remove data where atmospheric extinction was significant. The remaining data points are binned in sets of $50$, using consecutive \textsc{lstseq}, to a cadence of $320$ sidereal seconds. When a set of $50$ data points is incomplete, the existing points are binned and the number of raw data points used in the bin is recorded. Finally, the individual ${\sim} 15 \rm{~day}$ baselines are combined in sets of three months (`quarters') for each camera. We will refer to these calibrated and binned light curves as the quarterly data.

\section{Secondary calibration}
\label{sec:secondary_cal}

After removing the effects of the atmospheric extinction, transmission and intrapixel variations and binning the data to a cadence of $320$ sidereal seconds, residual systematic trends are observed in the quarterly light curves of individual stars. These residuals can be divided into two broad categories, first a daily repeating trend that depends primarily on the local sidereal time, which acts as a proxy for the position of the star on the CCD. Second, long term variations in the baseline. 

The daily trend is systematic in nature, caused by the variable shape of the stellar PSF with position on the CCD \citep{Talens2017a}, which changes the fraction of the stellar light within the photometric aperture and the amount of blended light from neighbouring stars in a way that is unique to each target. The long-term variations can be both physical and systematic in nature, with the main systematic long-term changes caused by an overestimation of the sky background when the moon is visible, producing an artificial dimming in the light curves that varies with the moon phase. Since we are interested in exoplanet transit signals we remove all long-term variations, indiscriminate of their origin. As an example of these trends Fig. \ref{fig:trends} shows the quarterly light curve of the known transiting exoplanet host star HD\,189733 as observed by the LPC camera during the third quarter of 2016 (2016Q3). The top panel shows the light curve as a function of the Julian date, revealing apparent dimming due to the Moon at ${\sim} 20$, ${\sim} 48$ and ${\sim} 76 \rm{~days}$. The bottom panel shows the light curve as a function of the local sidereal time, revealing the trend with CCD position resulting from changes in the PSF. The PSF effect is also responsible for the apparent scatter towards fainter magnitudes seen in the top panel at times ${<} 60 \rm{~days}$. This scatter is caused by the faint end of the PSF effect at sidereal times ${<} 18 \rm{~hours}$, which disappears later in the quarter as these sidereal times are no longer observed.

The methods presented below for removing these trends are applied to the quarterly light curves, before combining the different cameras and quarters for the transit search (Sect. \ref{sec:detection}). Applying these methods before the transit search is only valid because the signals of interest have low amplitudes and short durations, minimizing the impact of the signal on the removal of these trends. When the amplitude or the duration of the signal of interest are large, or when performing detailed modelling of any astrophysical signal present in the data, these methods should be run jointly with a model for the signal in question, which may be as simple as a phase-binned mean of the light curve \citep{Burggraaff2018}. Furthermore, when investigating long-term variability using MASCARA data the removal of long-term trends most be either turned off, or the long-term trends that were removed must be added back to the data. 

\begin{figure*}
  \centering
  \includegraphics[width=17cm]{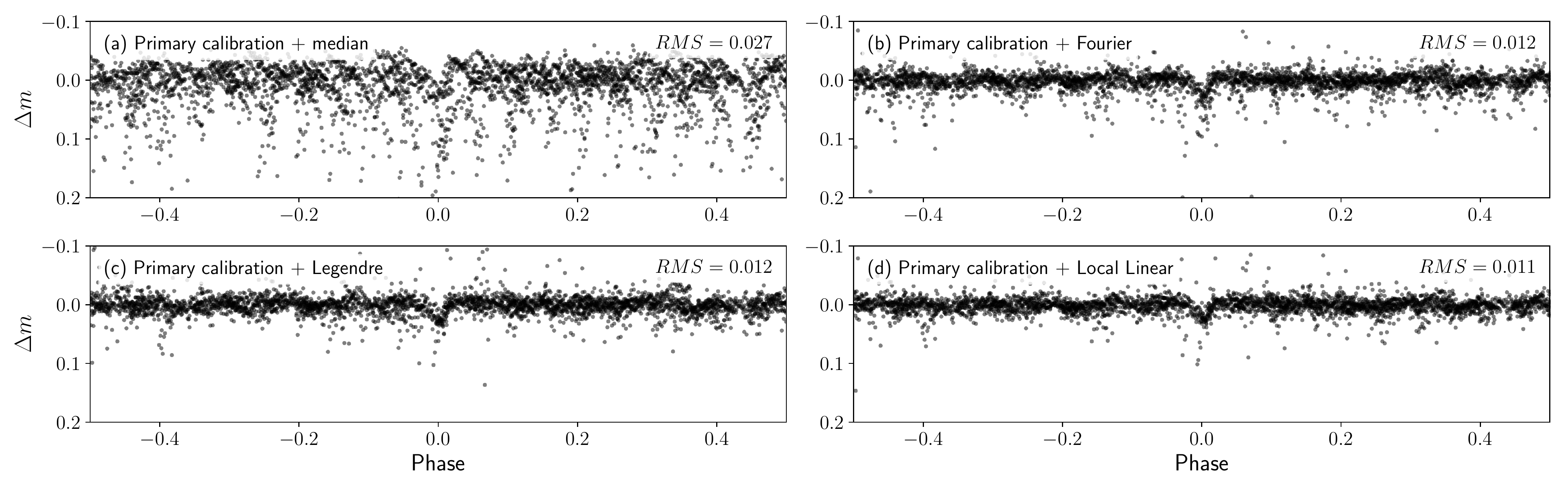}
  \caption{Light curve of HD\,189733 as observed by the LPC camera during 2016Q3. The data have been phase-folded to the orbital period of HD\,189733\,b. Panel (a) shows the light curve after primary calibration, with the median subtracted for clarity. Panels (b), (c) and (d) show the light curve for each of the secondary calibration methods, Fourier, Legendre and Local Linear respectively. The RMS scatter on the light curve is noted in the top-right of each panel.}
  \label{fig:trends_compare}
\end{figure*}

\subsection{Methods}

Three different methods for performing the secondary calibration were developed: Fourier, Legendre and Local Linear. We discuss each of these methods below.

\paragraph{Fourier} 

Guided by the daily repetition of the PSF effect, this method consists of the sum of two discrete Fourier series, one in the Julian date and one in the local sidereal time, with high frequencies removed to prevent over-fitting the data. The use of the Fourier method resulted in the successful detection of MASCARA-1\,b \citep{Talens2017b}; however, it is poorly motivated. The model does not account for global offsets and the number of degrees of freedom has to be artificially increased to account for the fact that the trends are not in fact periodic. This extra freedom comes from adding a sinusoid with half the lowest frequency to both series.

\paragraph{Legendre}

In order to improve on the shortcomings of the Fourier method a second method using Legendre polynomials was developed. Contrary to Fourier functions, Legendre polynomials ($P_n(x)$) form an orthogonal basis on a fixed interval, which matches better to the daily variations. A sum of two Legendre polynomials is used, one in the Julian date (\textsc{jd}) and one in the local sidereal time (\textsc{lst}). The \textsc{jd} and \textsc{lst} are scaled, and in the case of the \textsc{lst} wrapped, so that their values are optimally mapped onto the interval where the polynomials are orthogonal. In addition, a linear dependence on the sky background (\textsc{sky}) is included in the \textsc{lst} polynomial to correct for the effect of the Moon on the sky background level. In short, we model the magnitude as

\begin{align}
m_t = \sum_{m=0}^M (a_m + b_m \textsc{sky}_t) P_m(\textsc{lst}_t) + \sum_{n=1}^N c_n P_n(\textsc{jd}_t),
\label{eq:legendre}
\end{align}

where $a_m$, $b_m$ and $c_n$ are the polynomial coefficients. The polynomial orders $M$ and $N$ are determined from the baselines in \textsc{jd} and wrapped \textsc{lst} according to   

\begin{align}
M = \floor*{\frac{\textsc{lst}_{\rm{max}} - \textsc{lst}_{\rm{min}}}{s_{\textsc{lst}}}} \text{ and } N = \floor*{\frac{\textsc{jd}_{\rm{max}} - \textsc{jd}_{\rm{min}}}{s_{\textsc{jd}}}},
\end{align}

where $s_\textsc{lst} = 15 \rm{~min}$ and $s_\textsc{jd} = 5 \rm{~days}$ are the typical scales associated with the PSF effect and the long-term variations. We solve Eq. \ref{eq:legendre} iteratively, computing the residuals of the best-fit model at each iteration and excluding outliers from the next iteration at the $10\sigma$ level, where the median absolute deviation (\textsc{mad}) is used as a robust estimator of $\sigma$. The Legendre method was used in the detection of MASCARA-2\,b \citep{Talens2018a}, with $s_\textsc{jd} = 3 \rm{~days}$, without the linear \textsc{sky} dependence, and without the iterative scheme.

\paragraph{Local Linear}

The third method we developed does not depend on a particular choice of basis functions. Local Linear corrects the PSF effect by fitting a linear function of the \textsc{sky}, $x$ position and $y$ position to all measurements taken at the same \textsc{lst}, which are identified using the \textsc{lstidx} (see Sect. \ref{sec:introduction}), and removes the long-term variations using a weighted moving mean with a window $s_\textsc{jd} = 5 \rm{~days}$. The linear fit and moving mean are iteratively refined until convergence is reached. The linear function describing the PSF effect can be written as 

\begin{align}
m_t = a_n + b_n(x_t - \bar{x}_n) + c_n(y_t - \bar{y}_n) + d_n\textsc{sky}_t,
\label{eq:loclin}
\end{align}

\begin{figure*}
  \centering
  \includegraphics[width=17cm]{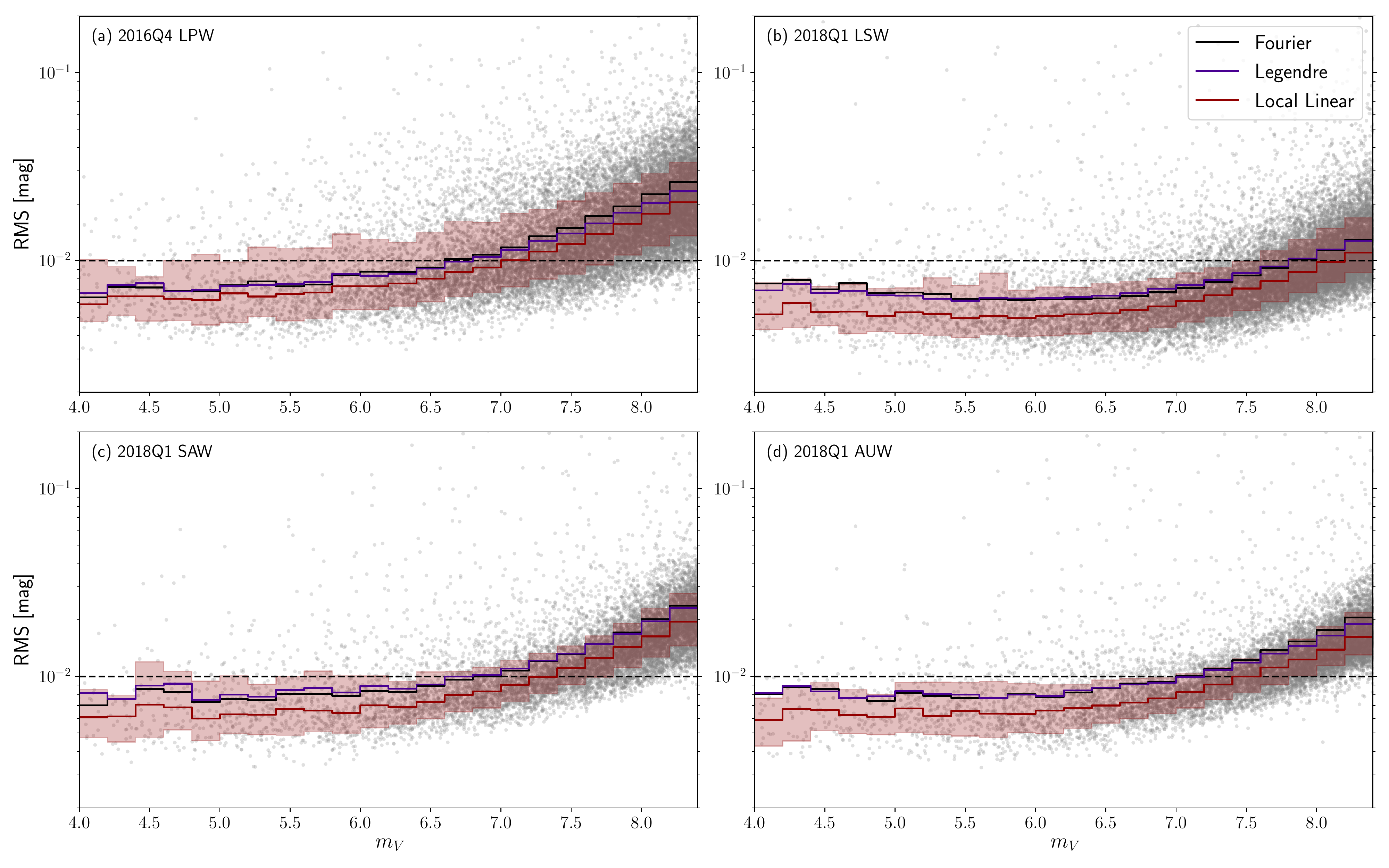}
  \caption{RMS scatter as a function of magnitude for stars observed by the west cameras of each station over the course of a quarter. The curves show the median RMS in bins of $0.2 \rm{~mag}$ for the three secondary calibration methods. For the Local Linear method the $1\sigma$ percentiles (coloured area), and the RMS scatter values of individual stars (grey points) are also indicated. The dashed black line shows the $10 \rm{~mmag}$ noise level. Only light curves with ${>} 500$ quarterly data points were used.}
  \label{fig:trends_rms}
\end{figure*}

where $n$ is short for the \textsc{lstidx}, $a_n$, $b_n$, $c_n$ and $d_n$ are the coefficients, $x_t$, $y_t$ and $\textsc{sky}_t$ are the CCD coordinates and sky background corresponding to the measurement and $\bar{x}_n$ and $\bar{y}_n$ are the average positions at each \textsc{lstidx}.

The inclusion of the $x$ and $y$ positions provides a marked improvement as they correct small changes in the measured fluxes resulting from astrometric drift, which is neglected in all the other calibrations. Unlike the Legendre method the Local Linear method does not include outlier rejection in its iterative scheme. The independent nature of the linear fits at different values of the \textsc{lstidx} makes it unsuitable, as outlier rejection might remove too many points at a particular \textsc{lstidx}.

\begin{figure*}
  \centering
  \includegraphics[width=17cm]{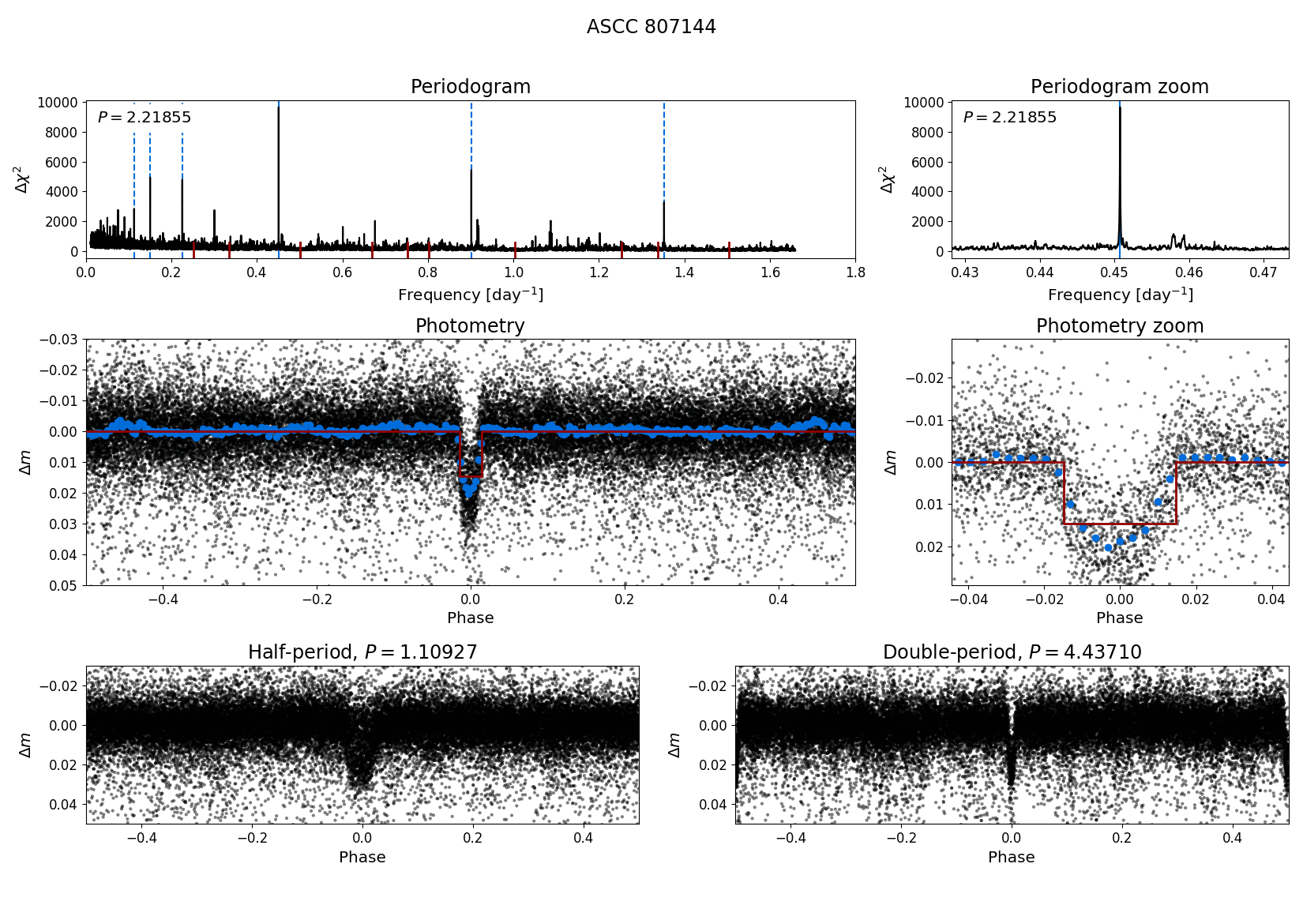}
  \caption{Diagnostic figure showing the results of a transit search run on three years of data from the La Palma station for the star HD\,189733. The top row shows the full BLS periodogram (left) and a zoom on the peak period (right). Both panel include the periodogram (black line), lines indicating the peak period and its harmonics (blue dashed lines) and lines indicating the sidereal period and its harmonics (red lines). The middle row shows the full phase-folded light curve (left) and a zoom on the transit (right), Both panels include the light curve (black dots), the phase-binned light curve (blue points) and the best-fit box model (red line). The bottom row shows the light curve phase-folded to half the peak period (left) and twice the peak period (right).}
  \label{fig:diagnostic}
\end{figure*}

\subsection{Comparison}

As an example of the differences between these methods Fig. \ref{fig:trends_compare} shows the quarterly light curve of HD\,189733, observed by the LPC camera during 2016Q3, as a function of the orbital phase of the hot Jupiter companion \citep[$P = 2.21857520 \rm{~days}$,][]{Baluev2015}. Panel (a) shows the data after the primary calibration and panels (b), (c) and (d) show a comparison between the three methods. The improvement from applying the secondary calibration methods to the light curve is evident, with over a factor two improvement in the RMS. The differences between the different methods are more subtle, with particularly the correction of outlier data points showing improvement going from Fourier to Legendre and finally to the Local Linear method. 

To quantify the merit of each of these methods Fig. \ref{fig:trends_rms} shows the RMS scatter as a function of magnitude for stars observed by the west cameras of each station during a specific quarter. We find that the Fourier and Legendre methods provide similar performance for all cameras at all magnitudes while the Local Linear method provides a marked improvement ranging from $\Delta\rm{RMS} \sim 1 \rm{~mmag}$ for the brightest stars to $\Delta\rm{RMS} \sim 3 \rm{~mmag}$ for the faintest stars. We also see that there are individual differences in performance between the different cameras with LPW reaching the $10 \rm{~mmag}$ RMS level at $m_V \sim 7.1$, SAW and AUW at $m_V \sim 7.3$, while AUE reaches this level at the fainter magnitude of $m_V \sim 8.1$. In addition to reaching $10 \rm{~mmag}$ RMS level at the brightest magnitude the LPW camera also shows the largest spread in achieved RMS as a function of magnitude. This difference between MASCARA La Palma and the southern stations is most likely a result of implementing a more robust method for obtaining the sky background on the southern stations. The relative performance of the three methods is a general feature observed in all MASCARA and bRing cameras, as are the individual performance differences between the cameras. 

\section{Transit detection}
\label{sec:detection}

\begin{table*}
\small
\centering
\caption{Results of varying the Legendre secondary calibration method. The default models for all secondary calibrations methods are shown in bold face.}
\begin{tabular}{l l l l l l}
 Name & $s_\textsc{jd}$ & $s_\textsc{lst}$ & \textsc{sky} term & Iterative & $N_{\rm{rec}}$ \\
 \hline
 & [days] & [min] & & & of 5000 \\
 \hline
 \textbf{Ref\_Fourier} & & & & & 1739\\
 \textbf{Ref\_LocLin} & & & & & 2267 \\
 Var\_sjd1p0\_NoIter & 1 & 15 & Yes & No & 885 \\
 Var\_sjd3p0\_NoIter\tablefootmark{a} & 3 & 15 & Yes & No & 1839 \\ 
 Var\_sjd3p0\_slst30\_NoIter & 3 & 30 & Yes & No & 1833 \\
 Var\_sjd3p0\_slst60\_NoIter & 3 & 60 & Yes & No & 1646 \\
 Var\_NoIter & 5 & 15 & Yes & No & 2003 \\
 Var\_slst30\_NoIter & 5 & 30 & Yes & No & 1987 \\
 Var\_slst60\_NoIter & 5 & 60 & Yes & No & 1800 \\
 Var\_sjd7p0\_NoIter & 7 & 15 & Yes & No & 1854 \\
 Var\_sjd7p0\_slst30\_NoIter & 7 & 30 & Yes & No & 1838 \\
 Var\_sjd7p0\_slst60\_NoIter & 7 & 60 & Yes & No & 1697 \\
 Var\_NoSky\_NoIter & 5 & 15 & No & No & 1854 \\
 Var\_slst30\_NoSky\_NoIter & 5 & 30 & No & No & 1838 \\
 Var\_slst60\_NoSky\_NoIter & 5 & 60 & No & No & 1697 \\
 Var\_Iter5sig & 5 & 15 & Yes & $5\sigma$ & 1995 \\
 \textbf{Ref\_Legendre} & 5 & 15 & Yes & $10\sigma$ & 2006 \\ 
 Var\_Iter15sig & 5 & 15 & Yes & $15\sigma$ & 2000 \\
 Var\_Iter20sig & 5 & 15 & Yes & $20\sigma$ & 2003 
\end{tabular}
\tablefoot{\tablefootmark{a}{Used in the discovery of MASCARA-2\,b \citep{Talens2018a}.}}
\label{tab:legendre_variations}
\end{table*}

In order to detect the periodic signals of transiting exoplanets we use the box least-squares algorithm \citep[BLS,][]{Kovacs2002} with the optimized parameter grid of \citet{Ofir2014}. The quarterly light curves are read and points binned from $\leq 45$ raw data points (out of a maximum of 50 per binning interval) are discarded before the secondary calibration is applied to individual cameras and quarters. The different cameras and quarters are then combined and the fully calibrated light curves are passed to the transit search algorithm where barycentric Julian dates are computed and outliers are rejected at the $3\sigma$ level, using the \textsc{mad} as a robust estimator of $\sigma$ and setting a lower limit of $50 \rm{~mmag}$ to prevent discarding transit signals. These data are then used to compute the BLS periodogram for each star.

After computing the BLS periodogram the strongest peak is identified and the parameters of the associated box-fit, namely orbital period, transit depth, transit duration and epoch are recorded. In addition, a number of diagnostic quantities, used in the literature to identify good transit candidates, are computed. We calculate the anti-transit ratio \citep{Burke2006}, the amplitude and the signal-to-noise ratio of ellipsoidal variations \citep{CollierCameron2006}, the signal detection efficiency \citep{Kovacs2002} and the signal-to-pink-noise \citep{Pont2006}. We further store the number of in-transit data points, the number of observed transits, the largest gap in the phase coverage relative to the transit duration, the fraction of points at $\rm{phase} < 0.5$ and the difference between the mean and median magnitude. Finally, the periodogram, best-fit parameters and diagnostic quantities are saved to disk.

Subsequently, diagnostic figures are made for all stars for which the strongest peak corresponds to a dimming of the star. As an example, Fig. \ref{fig:diagnostic} shows the diagnostic figure for HD\,189733 from a BLS run on three years of data from the La Palma station. It includes the BLS periodogram, the light curve phase-folded to the strongest peak, half the peak period and twice the peak period. All diagnostic quantities and figures are then inserted into a database that also contains the input catalogue \citep[ASCC,][]{Kharchenko2001,Kharchenko2009} estimates of stellar properties based on {\it Gaia} DR1 and DR2 \citep{GAIA2016a,GAIA2016b,GAIA2018} and links to SIMBAD \citep{Wenger2000} and the international variable star index \citep[VSX,][]{Watson2006}. The database contains a number of fixed queries tailored to finding candidate transiting exoplanets and allows for dynamic queries with constraints on multiple tables. Nominally, the fixed queries are used to select the most promising candidates before manually vetting them based on the diagnostic image and external information on known (nearby) variable stars. The status of each candidate is recorded in the database through the use of tags, indicating the current status, and user comments, recording the reasoning for the status and the history of the object. 

\section{Signal recovery tests}
\label{sec:recovery}

In order to characterize the performance of the primary and secondary calibration, as well as the performance of MASCARA as a transit survey, we carried out two sets of signal recovery tests. For these tests we injected transit signals in a subset of the light curves and ran the BLS algorithm to see how many signals we could recover successfully.

For each signal recovery test we drew a sample of 5000 stars for signal injection. To prevent this sample from being dominated by the fainter stars we set the probability of a star being selected proportional to $m_V^{-2}$, in addition we drew the sample without replacement, so no star is used twice. Subsequently, transit signals were injected with orbital periods ($P_{\rm{inj}}$) in the interval $[1,5) \rm{~days}$, epochs ($T_0$) in the interval $[0,P_{\rm{inj}}) \rm{~days}$, depths ($p^2 = (R_p/R_\star)^2$) from the set $\{0.005, 0.01, 0.02\}$, impact parameters ($b$) from the set $\{0.0, 0.5\}$ and stellar densities ($\rho_\star$) from the set $\{0.4, 0.9, 1.4\} \rm{~g~cm}^{-3}$. We used $\rho_\star$ as our fifth parameter, rather than the semi-major axis or transit duration, to ensure we injected realistic transits covering a range of spectral types. Transit light curves were computed from these parameters using \textsc{batman} \citep{Kreidberg2014}, including a linear limb darkening law with coefficient set to $0.6$, and injected into the light curves of the selected stars. Next we used the BLS algorithm to compute periodograms for all injected light curves and selected the period corresponding to the strongest peak, $P_{\rm{rec}}$. We counted a signal as recovered when

\begin{align}
\frac{|P_{\rm{inj}} - NP_{\rm{rec}}|}{P_{\rm{inj}}} < 10^{-3},
\end{align}

with $P_{\rm{rec}}$ the recovered period and $N=1,~2,~\frac{1}{2}\rm{~and~}\frac{1}{3}$, such that the harmonics are counted as successful recoveries.

\subsection{Performance of different calibration strategies}
\label{sec:recovery_calibration}

The goal of the first set of tests is to characterize the performance of the primary and secondary calibration, determining how the recovery rates change with variations on the calibration algorithms. As such we report in this section only the total number of recovered transit signals, independent of the transit parameters. We will discuss how the recovery rate depends on the transit parameters in Sect. \ref{sec:recovery_survey}, where we characterize the performance of MASCARA as a transit survey.

For this set of tests we injected signals in the raw $6.4 \rm{~s}$ light curves obtained during the fourth quarter of 2016 by the La Palma station (2016Q4). Before drawing the sample of 5000 stars for signal injection, we pre-selected those stars that had been observed for ${>} 300 \rm{~hours}$ across ${>} 60 \rm{~days}$, that is to say averaging at least five hours of data per night, to guarantee their light curves contained sufficient data for signal recovery to be possible. We used the same injected dataset in all recovery tests described in this section so that any change in the number of recovered signals, $N_{\rm{rec}}$, is due to changes in the calibration methods.

\begin{figure}
  \centering
  \includegraphics[width=8.5cm]{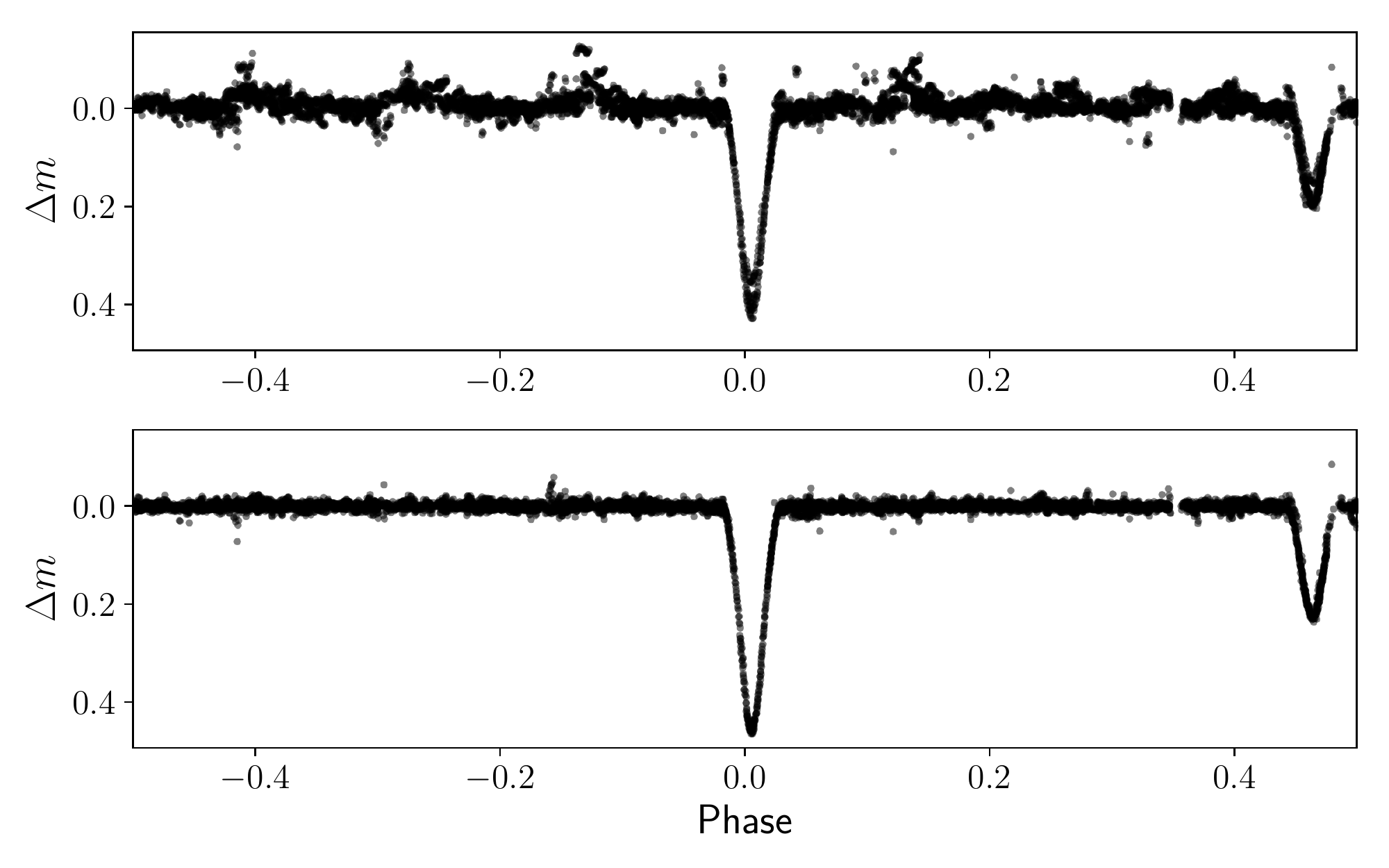}
  \caption{Light curve of the Algol-type eclipsing binary KW Hya, obtained by the La Silla station during 2017Q4 and 2018Q1, phase-folded to the orbital period of $P = 7.7499942 \rm{~days}$. The top panels shows the result of using the Legendre method without outlier rejection, while the bottom panel includes outlier rejection.}
  \label{fig:legendre_outlier}
\end{figure}

We started by keeping the primary calibration fixed and running variations on the Legendre secondary calibration method. For the Legendre method we varied the values of $s_{\textsc{jd}}$ and $s_{\textsc{lst}}$ which set the polynomial orders, investigated the effect of the linear \textsc{sky} dependence and the impact of excluding outliers when determining the best-fit parameters. Table \ref{tab:legendre_variations} lists the different variations and the number of recovered signals for each variation. Of the 5000 injected signals, around 2000 signals were recovered by the optimal variation of the Legendre method. Typically, $s_{\textsc{jd}} = 5 \rm{~days}$ is the optimal choice for removing the long-term variations with smaller values over-fitting the data (e.g. $s_{\textsc{jd}} = 1 \rm{~day}$) and larger values not providing enough freedom to correct for the long-term variations. For example, when $s_{\textsc{jd}} = 7 \rm{~days}$ the companion to HD\,189733 is not detected, as the intrinsic variability of the star is not sufficiently removed. For the PSF effect we find that the shortest value of $s_{\rm{lst}} = 15 \rm{~min}$ is preferred regardless of the value of $s_{\rm{jd}}$. There is also a preference for including the \textsc{sky} dependence with an increase in $N_{\rm{rec}}$ of $100-150$ signals when it is included. Finally, the inclusion of iterative outlier rejection appears to have little influence on the number of recovered signals. However, in practice we find that the iterative outlier rejection does have an effect on eclipsing binaries whose primary eclipses are relatively deep and short when compared to the orbital period, that is stars where the outlier rejection excludes the real signal from the fit. An example of such a star is shown in Fig. \ref{fig:legendre_outlier} with the top panel showing the light curve without outlier rejection and the bottom panel with outlier rejection. For this reason we keep the outlier rejection, and note that the Legendre secondary calibration method might be optimal for the detection of this type of binary. 

Next, we again kept the primary calibration fixed and ran variations on the Local Linear secondary calibration method. For the Local Linear method we tested the window size for the weighted running mean $s_{\rm{jd}}$, and investigated the effect of the positional and \textsc{sky} dependencies. Table \ref{tab:loclin_variations} lists the different variations and the number of recovered signals for each variation. Of the 5000 injected signals, 2267 signals are recovered by the optimal variation of the Local Linear method. As with the Legendre method, the Local Linear method performs optimally when the window of the long-term variations is set to $s_{\textsc{jd}} = 5 \rm{~days}$. The inclusion of both the positional and \textsc{sky} dependencies are also favoured with almost $200$ more detections when including the \textsc{sky} term and over $100$ more when including the positional terms. Most importantly, the best variation of the Local Linear method recovers over $250$ signals more than the best variation of the Legendre method.

Tables \ref{tab:legendre_variations} and \ref{tab:loclin_variations} show that of the three secondary calibration methods presented, the Local Linear method recovers the largest number of injected signals. However, comparing which of the 5000 signals are recovered by each method shows that a total of 237 signal are recovered by either or both of the Fourier and Legendre methods but not by the Local Linear method. Inspection of these signals revealed no apparent dependence on the injected transit parameters or stellar parameters, nor a particular systematic effect preventing their detection by the Local Linear method. As such we conclude that while the Local Linear is our preferred secondary calibration method, there is merit to continued use of the Fourier and Legendre methods as detectable signals may otherwise be missed. Furthermore, we continue to update existing methods and test new methods as our understanding of the data improves.

\begin{table}
\small
\centering
\caption{Results of varying the Local Linear secondary calibration method. The default models for all secondary calibrations methods are shown in bold face.}
\begin{tabular}{l l l l l}
 Name & $s_\textsc{jd}$ & $x,y$ terms & \textsc{sky} term & $N_{\rm{rec}}$ \\
 \hline
 & [days] & & & of 5000 \\
 \hline
 \textbf{Ref\_Fourier} & & & & 1739\\
 \textbf{Ref\_Legendre} & & & & 2006 \\
 Var\_sjd1p0 & 1 & Yes & Yes & 1495 \\
 Var\_sjd3p0 & 3 & Yes & Yes & 2234 \\
 \textbf{Ref\_LocLin} & 5 & Yes & Yes & 2267 \\
 Var\_sjd7p0 & 7 & Yes & Yes & 2206 \\
 Var\_sjd3p0\_NoPos & 3 & No & Yes & 2092 \\
 Var\_NoPos & 5 & No & Yes & 2126\\
 Var\_sjd7p0\_NoPos & 7 & No & Yes & 2079 \\
 Var\_sjd3p0\_NoSky & 3 & Yes & No & 2078 \\
 Var\_NoSky & 5 & Yes & No & 2075 \\
 Var\_sjd7p0\_NoSky & 7 & Yes & No & 1930 \\
\end{tabular}
\label{tab:loclin_variations}
\end{table}

\begin{figure}
  \centering
  \includegraphics[width=8.5cm]{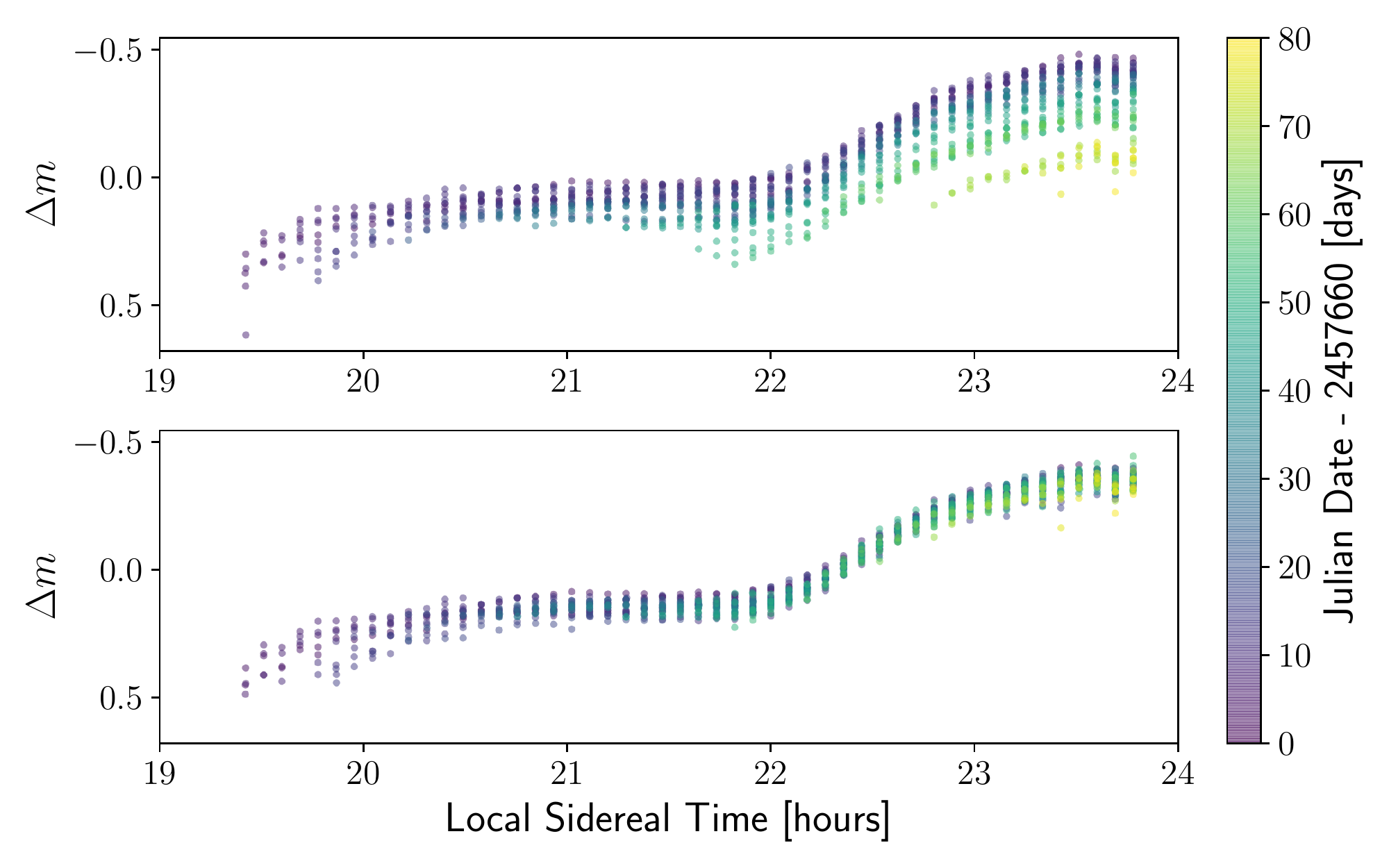}
  \caption{Light curve of HD\,205555, as observed by the LPC camera during 2016Q4, as a function of the local sidereal time and coloured by the Julian date of the observations. The top panels shows the data after primary calibration with the magnitudes, $m_i$, as free parameters, while the bottom panel shows the data after primary calibration with the $m_i = m_{V,i}$.}
  \label{fig:free_mag}
\end{figure}

Finally, we ran variations on the primary calibration, while using the optimal Local Linear secondary calibration method. We attempted to increase the \textsc{ha} resolution of the intrapixel amplitude maps from $320 \rm{~s}$ to $128 \rm{~s}$, investigated the effect of leaving the magnitudes $m_i$ as free parameters and tried to use a polar grid with declination rings of a fixed area, avoiding the excess of rings containing ${<}50$ stars for cameras observing the poles (see Fig. \ref{fig:hist}) at the cost of resolution near the poles. We did not vary the total number of rings in the polar grid or the resolution of the HEALPix grid as they are strongly motivated by the number of available stars. The same goes for the \textsc{ha} resolution of the transmission map which is motivated by the observational cadence. Table \ref{tab:primary_variations} lists the different variations and the number of recovered signals for each variation. We find that increasing the \textsc{ha} resolution of the intrapixel amplitudes slightly decreases the number of recovered signals, as does using equal-area declination rings in the polar grid. When we leave the magnitudes as a free parameter however $N_{\rm{rec}}$ increase by $18$ signals, and if we combine the free magnitudes with equal-area declination rings the recovery rate increase by $49$ signals compared to the reference model. We nevertheless prefer to keep the magnitudes fixed, as this ensures a fixed reference level across timescales longer than the ${\sim} 15$ day calibration baselines. Figure \ref{fig:free_mag} shows the light curve of HD\,205555 as a function of the \textsc{lst} and coloured by the Julian date, the top panel shows the result with the $m_i$ as free parameters and the bottom panel with the $m_i = m_{V,i}$. In the case of the free $m_i$ the normalization of the data changes with the \textsc{lst} coverage. During the first calibration baseline of the quarter we observed the full \textsc{lst} range shown and thus the full PSF effect while during subsequent baselines the observed \textsc{lst} values and the associated section of the PSF effect changes, shifting the corresponding sections of data. When we keep the $m_i$ fixed we find that this shift does not occur as we are using the same mean magnitudes across the quarter. In the case of HD\,205555 the shifts in $m_i$ are caused by the changing \textsc{lst} coverage, but intrinsic variability has the same effect even in the absence of strong \textsc{lst} variations. For example, an RR Lyrae type variable will be observed with different phase coverage during every ${\sim} 15$ day baseline, resulting in similar shifts when leaving the $m_i$ as free parameters. 

\begin{table}
\small
\centering
\caption{Results of varying the primary calibration. All recovery test were run with the Local Linear secondary calibration method.}
\begin{tabular}{l l l l l}
 Name & $m_i$ & Polar rings & Index $l$ & $N_{\rm{rec}}$ \\
 \hline
 & & & & of 5000 \\
 \hline
 \textbf{Ref\_Primary} & $m_{V,i}$ & Fixed width & $320 \rm{~s}$ & 2267\\
 Var\_Nl675 & $m_{V,i}$ & Fixed width & $128 \rm{~s}$ & 2242\\
 Var\_EqArea & $m_{V,i}$ & Fixed area & $320 \rm{~s}$ & 2255\\
 Var\_mFree & $m_i$ & Fixed width & $320 \rm{~s}$ & 2285\\
 Var\_mFree\_EqArea & $m_i$ & Fixed area & $320 \rm{~s}$ & 2316\\
\end{tabular}
\label{tab:primary_variations}
\end{table}

\subsection{Survey performance}
\label{sec:recovery_survey}

The goal of the second set of tests is to characterize the performance of MASCARA as a transit survey, determining how the recovery rates depend on stellar and transit parameters. For these test we used one year (2016) and three years (2015-2017) of observations obtained by the La Palma station. For these tests a sample of 5000 stars was drawn without any pre-selection on the amount of data obtained for each star, and the signals were injected in the light curves after the primary calibration. Injecting the signals after the primary calibration is not expected to have a major impact on the result, as the presence of injected signals on the primary calibration was found to be negligible for the recovery tests presented in Sect. \ref{sec:recovery_calibration}, with ${\sim}95.5\%$ of the correction terms changed by ${<} 1 \rm{~mmag}$. In order to improve the statistics each light curve was duplicated 11 times, with different transit signals injected in the first ten duplicates and the 11th acting as a reference. For the test using three years of data we extend the range of $P_{\rm{inj}}$ to $[1,10) \rm{~days}$. We used the optimal Local Linear secondary calibration method for both the one year and three years of data.

\begin{figure*}
  \centering
  \includegraphics[width=17cm]{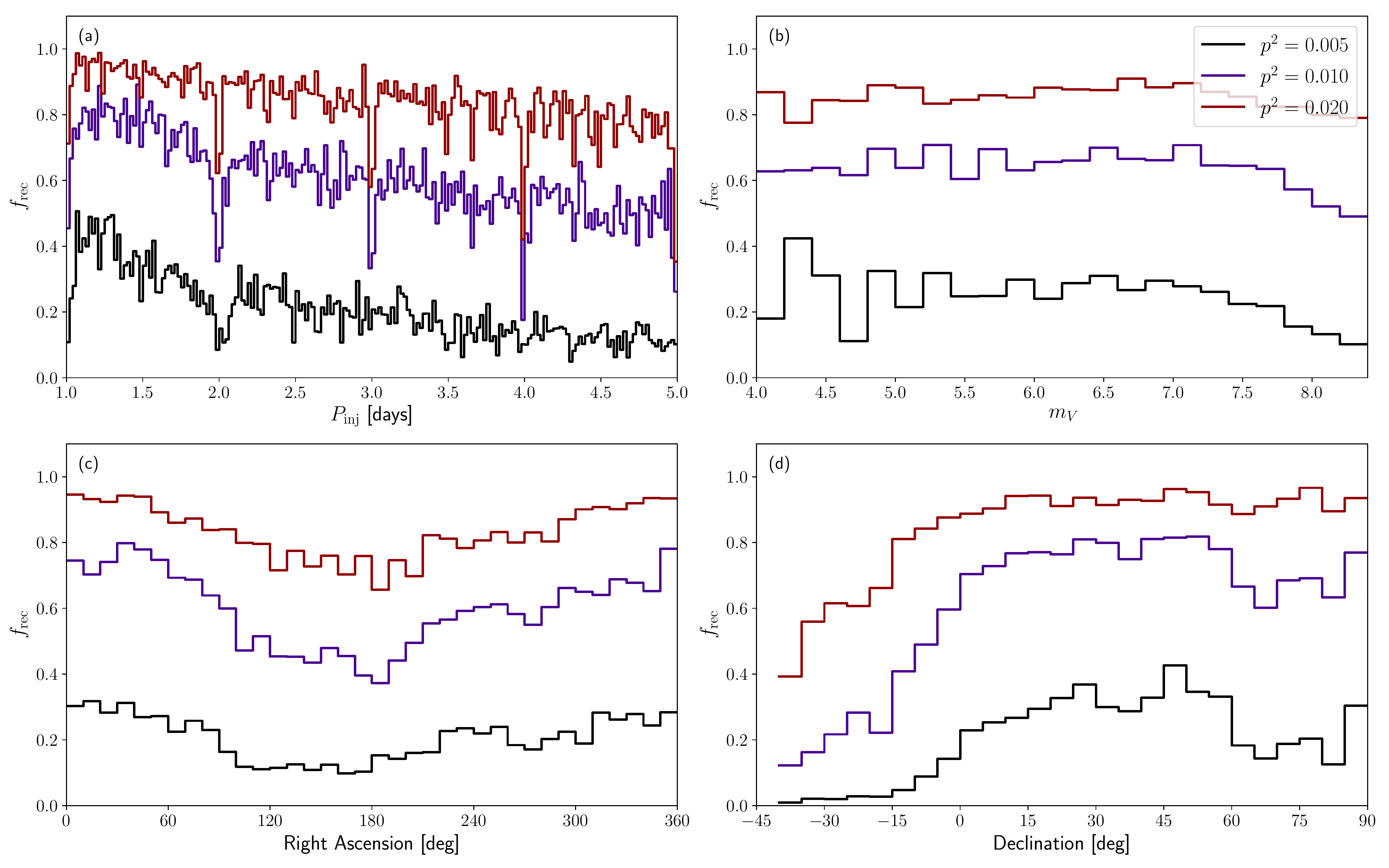}
  \caption{Fraction of transit signals recovered from one year of data from the La Palma station. The curves show the fraction of recovered signals for the different transit depths. The panels show the recovery rates as a function of different parameters: \textit{(a)} as a function of the injected period in 192 bins of $0.5$ hours, \textit{(b)} as a function of the visual magnitude in 22 bins of $0.2$ magnitudes, \textit{(c)} as a function of the right ascension in 36 bins of $10\degr$ and \textit{(d)} as a function of the declination in 27 bins of $5\degr$.}
  \label{fig:frec_Q5Q8LPa0loclin}
\end{figure*}

After running the signal recovery test on one year of data we found that seven stars were rejected by the BLS algorithm because insufficient data was available. We further removed 119 stars with $m_V < 4$, focusing on the range $4 < m_V < 8.4$ where MASCARA was designed for optimal performance but we note that the input catalogue does contain brighter stars, of the 48,740 remaining signals $54.8 \%$ are recovered overall. Table \ref{tab:frec_Q5Q8LPa0loclin} lists the recovery fractions for a number of different subsets of the sample and Fig. \ref{fig:frec_Q5Q8LPa0loclin} shows the recovery fractions as a function of $P_{\rm{inj}}$, $m_v$, $\alpha$ and $\delta$ for the different transit depths $p^2$. 

The recovery fraction depends only weakly on the values of $b$ and $\rho_\star$, ranging from ${\sim} 50$ to ${\sim} 60 \%$, with higher values of both parameters resulting in lower recovery fractions. This dependence is expected as higher values of $b$ and $\rho_\star$ result in shorter transits for fixed orbital periods, reducing the amount of in-transit data and increasing the total amount of data needed to obtain a detection. The recovery fraction depends strongly on $p^2$, recovering $84$, $60.5$ and $20.7 \%$ for the different depths, and showing a steep drop in recovered signals going from $p^2=0.01$ to $p^2=0.005$. The dependence on the injected period is shown in panel (a) of Fig. \ref{fig:frec_Q5Q8LPa0loclin} for each value of $p^2$. The recovery fraction decreases as a function of the injected period, as fewer transits are observed on a fixed baseline, and strong dips are present at integer multiples of one day, where ground-based surveys are less sensitive due to gaps in the coverage. 

The dependence on stellar magnitude, right ascension and declination are shown in panels (b), (c) and (d) of  Fig. \ref{fig:frec_Q5Q8LPa0loclin} for each value of $p^2$. As a function of $m_V$ the recovery rate is relatively flat for $m_V < 7.5$, before dropping off towards fainter magnitudes, qualitatively matching the achieved RMS as a function of magnitude (see Fig. \ref{fig:trends_rms}). We find a sinusoidal behaviour with $\alpha$, which can be explained by the changes in the length of the night across the year as stars observed in winter will have more data available. Finally, as a function of the declination we find that the recovery rate peaks between $0\degr < \delta < 60\degr$ with a small drop at $\delta > 60\degr$ and a steep drop-off at $\delta < 0\degr$. These features can be explained by the on-sky coverage of the La Palma station \citep[Fig. 2]{Talens2017a}, at $\delta > 60\degr$ the sky is observed by only the LPN camera while between $-20\degr < \delta < 60\degr$ we cover the sky with three or four cameras, and with significant overlap between the cameras. The recovery fraction drops at $\delta \sim 0 \degr$ rather that $\delta \sim -20\degr$ because the width of the hour angle window as a function of the declination becomes the limiting factor on the amount of data obtained for each star. The \textsc{ha} coverage is ${\sim} 8 \rm{~h}$ at $\delta \sim 0\degr$, comparable to the length of the shortest night, and shrinks to ${\sim} 5 \rm{~h}$ at $\delta \sim -20 \degr$. The area below $\delta = 0\degr$ accounts for a large fraction of the total observed area, so we also report recovery rates for $\delta > 0 \degr$ and $\delta < 0 \degr$ as $65.4$ and $34.8 \%$, respectively. 

\begin{figure*}
  \centering
  \includegraphics[width=17cm]{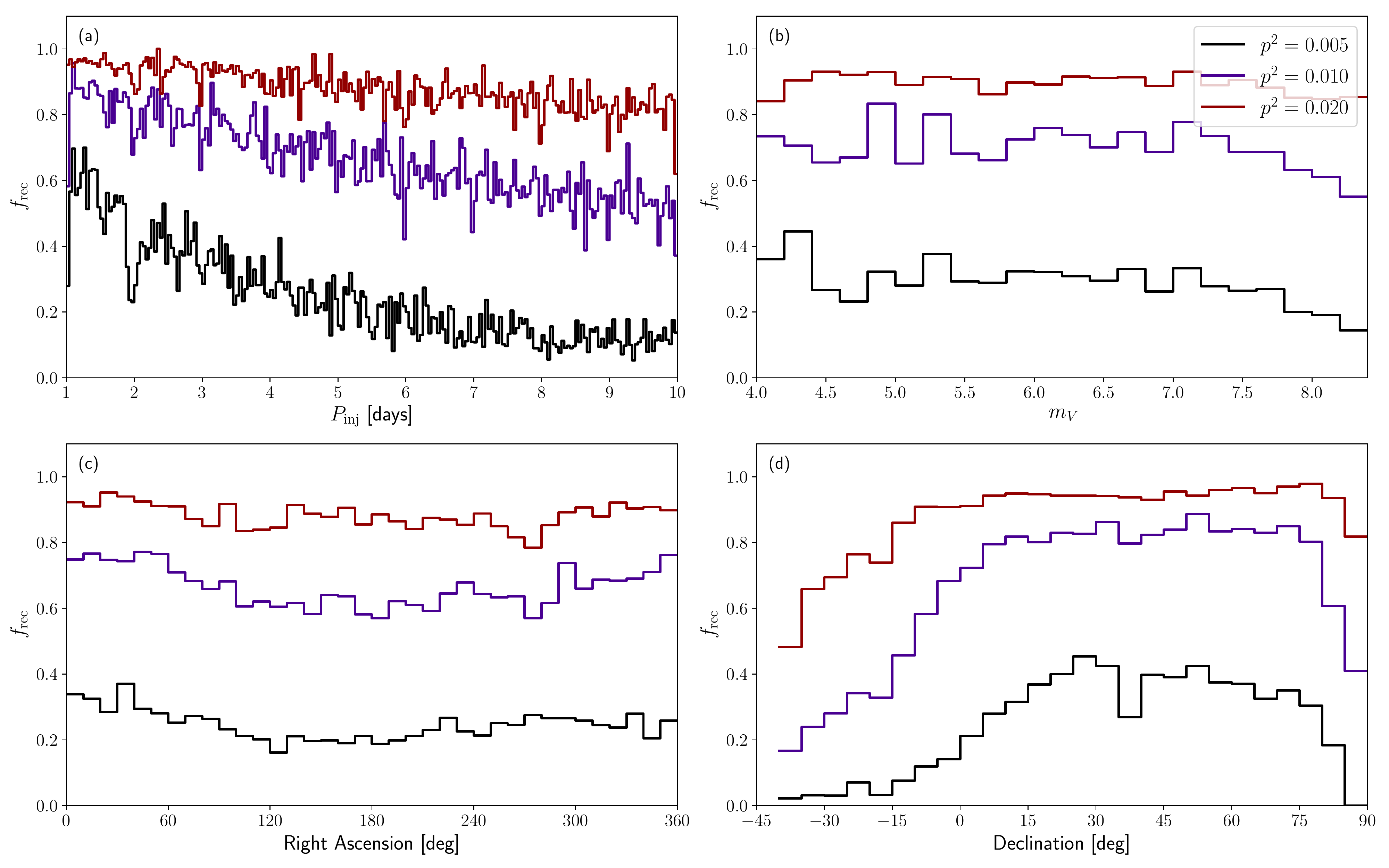}
  \caption{Fraction of transit signals recovered from three years of data from the La Palma station. Panels and curves are the same as in Fig. \ref{fig:frec_Q5Q8LPa0loclin} with the exception of panel \textit{(a)}, which now shows the data in 216 bins of $1$ hour.}
  \label{fig:frec_Q1Q12LPa0loclin}
\end{figure*}

After running the signal recovery test on three years of data we found three stars were rejected by the BLS algorithm because insufficient data was available, and we removed 125 stars with $m_V < 4$, leaving 48,720 signals. The results are listed in Table \ref{tab:frec_Q5Q8LPa0loclin} and shown if Fig. \ref{fig:frec_Q1Q12LPa0loclin}. 

With three years of MASCARA data we obtain slightly better performance out to ten day orbital periods than we did out to five day periods with only one year of data. We also find similar dependencies on the transit and stellar parameters, and a slightly improved recovery rate for periods ${<}5 \rm{~days}$, particularly for the shallowest transits ($p^2 = 0.005$). From panels (b) and (c) of Fig. \ref{fig:frec_Q1Q12LPa0loclin} we see that for the deepest transits ($p^2=0.02$) the dependencies on $m_V$ and $\alpha$ show less structure than those of shallower transits, suggesting that for the deeper transits we are no longer limited by the RMS or changes in the data volume with time of year, though from panel (d) we see that changes in the data volume with declination remain a limiting factor. The only other major difference is a decrease in the number of signals recovered in the declination bin closes to the pole for $p^2=0.01$ and $p^2=0.005$, however this could be a consequence of the small area of this bin, which reduces the number of available stars and increases the uncertainty on the result. In fact, for the three year recovery test only five distinct stars were selected in this declination bin. Considering only the region of parameter space where MASCARA La Palma performs best we find that with three years of data we recover a total $93 \%$ of signals with $1 < P < 5 \rm{~days}$ and depths $\geq 1\%$ at $\delta > 0\degr$, versus $84 \%$ in one year of data.  

\begin{table}
\centering
\caption{Results of signal recovery tests on data from the La Palma station, for one year of data with periods between one and five days and three years of data with periods between one and ten days.}
\begin{tabular}{l l l}
 Selection & $f_{\rm{rec}}$ [\%] & $f_{\rm{rec}}$ [\%] \\ 
 \hline
 & 1 year & 3 years \\
 \hline
 All & 54.8 & 59.6\\
 $\delta > 0\degr$ & 65.4 & 70.1\\
 $\delta < 0\degr$ & 35.8 & 40.8\\
 $p^2 = 0.005$ & 20.7 & 24.7 \\
 $p^2 = 0.01$ & 60.5 & 66.7 \\
 $p^2 = 0.02$ & 84.0 & 88.1 \\
 $b = 0.0$ & 57.7 & 62.7 \\
 $b = 0.5$ & 51.8 & 56.4 \\
 $\rho_\star = 0.4 \rm{~g~cm}^{-3}$ & 59.5 & 64.1 \\
 $\rho_\star = 0.9 \rm{~g~cm}^{-3}$ & 54.1 & 59.0 \\
 $\rho_\star = 1.4 \rm{~g~cm}^{-3}$ & 50.9 & 55.6 \
\end{tabular}
\label{tab:frec_Q5Q8LPa0loclin}
\end{table}

\section{The A-star occurrence rate}
\label{sec:occurrence}

Using the results of the signal recovery analysis, and the hot Jupiters detected by MASCARA La Palma, it is possible to place a first constraint on the occurrence of hot Jupiter around A-type stars. From radial velocity (RV) surveys the occurrence rate of hot Jupiters around FGK-stars is estimated to to be ${\sim} 1\%$ \citep[e.g.][]{Mayor2011,Wright2012}, while transit surveys such as \emph{Kepler} suggest a lower occurrence rate of ${\sim} 0.4\%$ \citep[e.g.][]{Howard2012,Fressin2013}. Furthermore, RV surveys of evolved stars indicate that the occurrence rate of massive planets increases with stellar mass \citep[e.g.][]{Reffert2015,Ghezzi2018}. The occurrence rate of hot Jupiters around A-stars has not been constrained, because A-stars are typically excluded from RV surveys, and difficult to follow-up at the fainter magnitudes probed by most transit surveys. Within the magnitude range observed by MASCARA ($4 < m_V < 8.4$) four planets have been discovered transiting A-stars, WASP-33, MASCARA-1, MASCARA-2 and KELT-9 \citep{CollierCameron2010,Talens2017b,Talens2018a,Lund2017,Gaudi2017} all of which are detected in the three year dataset from MASCARA La Palma. All of these stars are located in the northern hemisphere ($\delta > 0\degr$) and the transits have depths ranging from $0.7$ to $1.4\%$. The ASCC catalogue contains 7,320 A-stars in the northern hemisphere. From this we can make a simple estimate of the occurrence rate of hot Jupiters around A-stars, correcting for the geometric transit probability \citep[$R_\star/a$,][]{Borucki1984} and the completeness of MASCARA in the northern hemisphere ($70.1 \%$, see Table \ref{tab:frec_Q5Q8LPa0loclin}). We find an occurrence rate of $0.4 \%$ for hot Jupiters around A-stars, consistent with results for FGK-stars from other transits surveys. This is nevertheless a preliminary estimate, and  should be considered a lower limit, as the sample size is small and incomplete. Several MASCARA candidates still await follow-up in the northern hemisphere and observations in the southern hemisphere are just beginning. Furthermore, currently only the significantly bloated hot Jupiters are found, as a $1~R_{\rm{J}}$ planet around a $2~R_{\odot}$ A-star would produce a $0.25 \%$ transit, and may be missed.

\section{Summary and outlook}
\label{sec:conclusions}

In this paper we described the methods used to process the raw photometry obtained by MASCARA and bRing for the purpose of detecting transiting exoplanets. For the primary calibration we use a modified version of the coarse decorrelation algorithm \citep{CollierCameron2006} to remove instrumental systematics, next we apply a secondary calibration method designed to remove residual systematics from individual stars and remove long-term variations, which are not of interest in the context of transit detection. Finally, we search these data for transit signals using the Box Least-Squares algorithm and place the results in a database which we use to select and vet promising candidates for follow-up observations. 

In order to assess the quality of the calibrated data and the ability of MASCARA to detect transiting exoplanets we computed the RMS scatter as a function of magnitude, demonstrating an RMS of 10 mmag is obtained at $m_V \sim 7.5$, depending on the camera considered. In addition, we performed signal recovery tests, a more powerful diagnostic of the data quality in the context of transit detection, demonstrating that with one year of data obtained by the La Palma station we can recover $54.8 \%$ of signals with periods between one and five days. With this test we identified two main causes for missed detections. First, the quality of the photometry, with $84$, $60.5$ and $20.7 \%$ of $2$, $1$ and $0.5 \%$ depth transits recovered, respectively. Second, the data volume obtained at different declinations, recovering $65.4 \%$ at $\delta > 0 \degr$ versus $35.8\%$ at $\delta < 0 \degr$. Finally, we demonstrated that with the full three years of data available for the La Palma station we can obtain similar recovery rates on signals with periods between one and ten days, and recover $93 \%$ of transits of $1\%$ or deeper with $1 < P < 5 \rm{~days}$ at $\delta > 0\degr$.

Over the course of 2017 the MASCARA station at La Silla and the two bRing stations have begun observing the southern sky. At the time of writing we have not yet obtained a full years worth of data from MASCARA La Silla, preventing us from performing a comparable signal recovery test for the southern hemisphere. However, based on the RMS curves presented in Fig. \ref{fig:trends_rms} we expect to obtain at least similar, and probably improved, performance with MASCARA La Silla. The La Silla station observes the sky between $-90\degr < \delta < 40\degr$, providing the region with $-40\degr < \delta < 0\degr$ with coverage comparable to the that provided by MASCARA La Palma between $0\degr < \delta < 40\degr$, filling in the gap where the recovery rates from the La Palma station are low. Furthermore, at the declinations where the MASCARA La Silla FoV overlaps with the bRing FoV, at $\delta \lesssim -30 \degr$, the near continuous temporal coverage obtained reduces the baseline needed for reliable transit detections at a given orbital period. Using data from the southern stations a number of transit candidates have already been identified and follow-up efforts have started.  

Using the results from our signal recovery tests, and the hot Jupiters detected by MASCARA La Palma, we placed a lower limit of $0.4 \%$ on the occurrence rate of hot Jupiters around A-type stars. This number is consistent with previous results for the occurrence rates of hot Jupiters around FGK stars from RV \citep[e.g.][]{Mayor2011,Wright2012} and transit surveys \citep[e.g.][]{Howard2012,Fressin2013}. We expect to be able to improve our constraint on the occurrence rate in the future, as additional candidate planets still await verification in the north and south, but caution that ${\sim} 1 ~R_{\rm{J}}$ hot Jupiters orbiting A stars are beyond the photometric precision of MASCARA and bRing. 

Our ability to blindly find signals with periods as long as $10 ~\rm{days}$ demonstrates the possibilities of combining MASCARA data with that of the TESS satellite \citep{Ricker2015}, which was launched in April 2018. TESS will obtain photometry for the same stars observed by MASCARA at higher precision but on much shorter baselines, $27$ days over most of the sky with longer baselines closer to the ecliptic poles. As a result TESS might only observe one transit event for periods between $13.5$ and $27 \rm{~days}$, placing constraints on the transit shape and epoch, but not the period. However, this information can be used to run specialized transit searches on the MASCARA data of these stars, boosting the signal by utilizing the transit shape and epoch from TESS and by running the search jointly with the secondary calibration methods. In this way, we expect to be able to place constraints on the orbital periods of some of the TESS single transit candidates, making potential follow-up of these targets less time intensive. 

\begin{acknowledgements}
I.S. acknowledges support from a NWO VICI grant (639.043.107). This project has received funding from the European Research Council (ERC) under the European Union's Horizon 2020 research and innovation programme (grant agreement nr. 694513). E.E.M. and S.N.M. acknowledge support from the NASA NExSS programme. S.N.M. is a U.S. Department of Defense SMART scholar sponsored by the U.S. Navy through SSC-LANT. Part of this research was conducted at Jet Propulsion Laboratory, California Institute of Technology, under a contract with the National Aeronautics and Space Administration. Construction of the bRing observatory at Siding Springs, Australia would not be possible without a University of Rochester University Research Award, help from Mike Culver and Rich Sarkis (UR), and generous donations of time, services, and materials from Joe and Debbie Bonvissuto of Freight Expediters, Michael Akkaoui and his team at Tanury Industries, Robert Harris and Michael Fay at BCI, Koch Division, Mark Paup, Dave Mellon, and Ray Miller and the Zippo Tool Room. This work has made use of data from the European Space Agency (ESA) mission {\it Gaia} (\url{https://www.cosmos.esa.int/gaia}), processed by the {\it Gaia} Data Processing and Analysis Consortium (DPAC,\url{https://www.cosmos.esa.int/web/gaia/dpac/consortium}). Funding for the DPAC has been provided by national institutions, in particular the institutions participating in the {\it Gaia} Multilateral Agreement. This research has made use of the SIMBAD database, operated at CDS, Strasbourg, France. This research has made use of the International Variable Star Index (VSX) database, operated at AAVSO, Cambridge, Massachusetts, USA. We have benefited greatly from the publicly available programming language {\sc Python}, including the {\sc numpy, matplotlib, astropy, scipy} and {\sc h5py} packages.
\end{acknowledgements}

\bibliographystyle{aa}
\bibliography{mascara.bib}

\appendix
\section{Details on solving Eqs. \ref{eq:loglike_spatial} and \ref{eq:loglike_temporal}}
\label{app:solution}

\subsection{Spatial}

Since the index $n$ solely depends on quantities related to the star $i$ Eq. \ref{eq:loglike_spatial} can be written as

\begin{align}
\ln L = -\frac{1}{2} \sum_n \sum_{i\in n,t} \bigg[\ln(2\pi\bar{\sigma}_{it}^2) + \frac{(\bar{m}_{it} - T_{nk} - f(x_{it}, y_{it}))^2}{\bar{\sigma}_{it}^2}\bigg],
\end{align}

and each ring $n$ can be solved independently. Setting the derivatives to zero results in:

\begin{align}
\frac{\partial}{\partial T_{nk}} \ln L_n = 0 = & \sum_{i,t \in n, k} \frac{\bar{m}_{it} - T_{nk} - f(x_{it}, y_{it})}{\bar{\sigma}_{it}^2} \label{eq:trans_sol}\\
\frac{\partial}{\partial a_{nl}} \ln L_n = 0 = & \sum_{i,t \in n, l} \frac{\bar{m}_{it} - T_{nk} - f(x_{it}, y_{it})}{\bar{\sigma}_{it}^2}. \label{eq:ipx_sol}
\end{align}

These equations are solved iteratively for $T_{nk}$ and the amplitudes $a_{nl}$, $b_{nl}$, $c_{nl}$, $d_{nl}$. First $T_{nk}$ is obtained from Eq. \ref{eq:trans_sol} while keeping the amplitudes fixed. Then the amplitudes are obtained simultaneously by solving Eq. \ref{eq:ipx_sol} using a linear least-squares solver while keeping $T_{nk}$ fixed. These steps are then repeated until convergence is reached. 

\subsection{Temporal}

Since the index $q$ solely depends on quantities related to the star $i$ Eq. \ref{eq:loglike_temporal} can be written as

\begin{align}
\ln L = -\frac{1}{2} \sum_q \sum_{i\in q,t} \bigg[\ln(2\pi(\sigma_{it}^2 + \sigma_i^2 + \sigma_{qt}^2)) + \frac{(\tilde{m}_{it} - m_i - c_{qt})^2}{\sigma_{it}^2 + \sigma_i^2 + \sigma_{qt}^2}\bigg],
\end{align}

and each patch $q$ can be solved independently. Taking derivatives with respect to the variables $m_i$, $\sigma_i$ and $c_{qt}$, $\sigma_{qt}$ and setting them to zero gives:

\begin{align}
\frac{\partial}{\partial m_i} \ln L_q = 0 = & \sum_t \frac{\tilde{m}_{it} - m_i - c_{qt}}{\sigma_{it}^2 + \sigma_i^2 + \sigma_{qt}^2} \label{eq:mi_sol}\\
\frac{\partial}{\partial \sigma_i} \ln L_q = 0 = & \sum_t \bigg[-\frac{1}{\sigma_{it}^2 + \sigma_i^2 + \sigma_{qt}^2} + \frac{(\tilde{m}_{it} - m_i - c_{qt})^2}{(\sigma_{it}^2 + \sigma_i^2 + \sigma_{qt}^2)^2}\bigg] \label{eq:sigi_sol}\\
\frac{\partial}{\partial c_{qt}} \ln L_q = 0 = & \sum_{i \in q} \frac{\tilde{m}_{it} - m_i - c_{qt}}{\sigma_{it}^2 + \sigma_i^2 + \sigma_{qt}^2} \label{eq:cqt_sol}\\
\frac{\partial}{\partial \sigma_{qt}} \ln L_q = 0 = & \sum_{i \in q} \bigg[-\frac{1}{\sigma_{it}^2 + \sigma_i^2 + \sigma_{qt}^2} + \frac{(\tilde{m}_{it} - m_i - c_{qt})^2}{(\sigma_{it}^2 + \sigma_i^2 + \sigma_{qt}^2)^2}\bigg]. \label{eq:sigqt_sol}
\end{align}

These equations are then solved iteratively for $\sigma_i$ (fixing $m_i = m_{V,i}$ as described in Sect. \ref{sec:primary_cal}) and $c_{qt}$, $\sigma_{qt}$. First $\sigma_i$ is obtained by using a bisector method to solve Eq. \ref{eq:sigi_sol} while keeping $c_{qt}$ and $\sigma_{qt}$ fixed. Then $c_{qt}$ and $\sigma_{qt}$ are jointly solved using a bisector method along $\sigma_{qt}$, first solving Eq. \ref{eq:cqt_sol} for $c_{qt}$ before evaluating Eq. \ref{eq:sigqt_sol} for the next bisection, all while keeping $\sigma_i$ fixed. For the bisector method we consider an interval of $[0,2)$ for both $\sigma_i$ and $\sigma_{qt}$, and iterate until convergence is reached.

\section{Correction maps}
\label{app:extra_maps}

\begin{figure*}
  \centering
  \includegraphics[width=8.5cm]{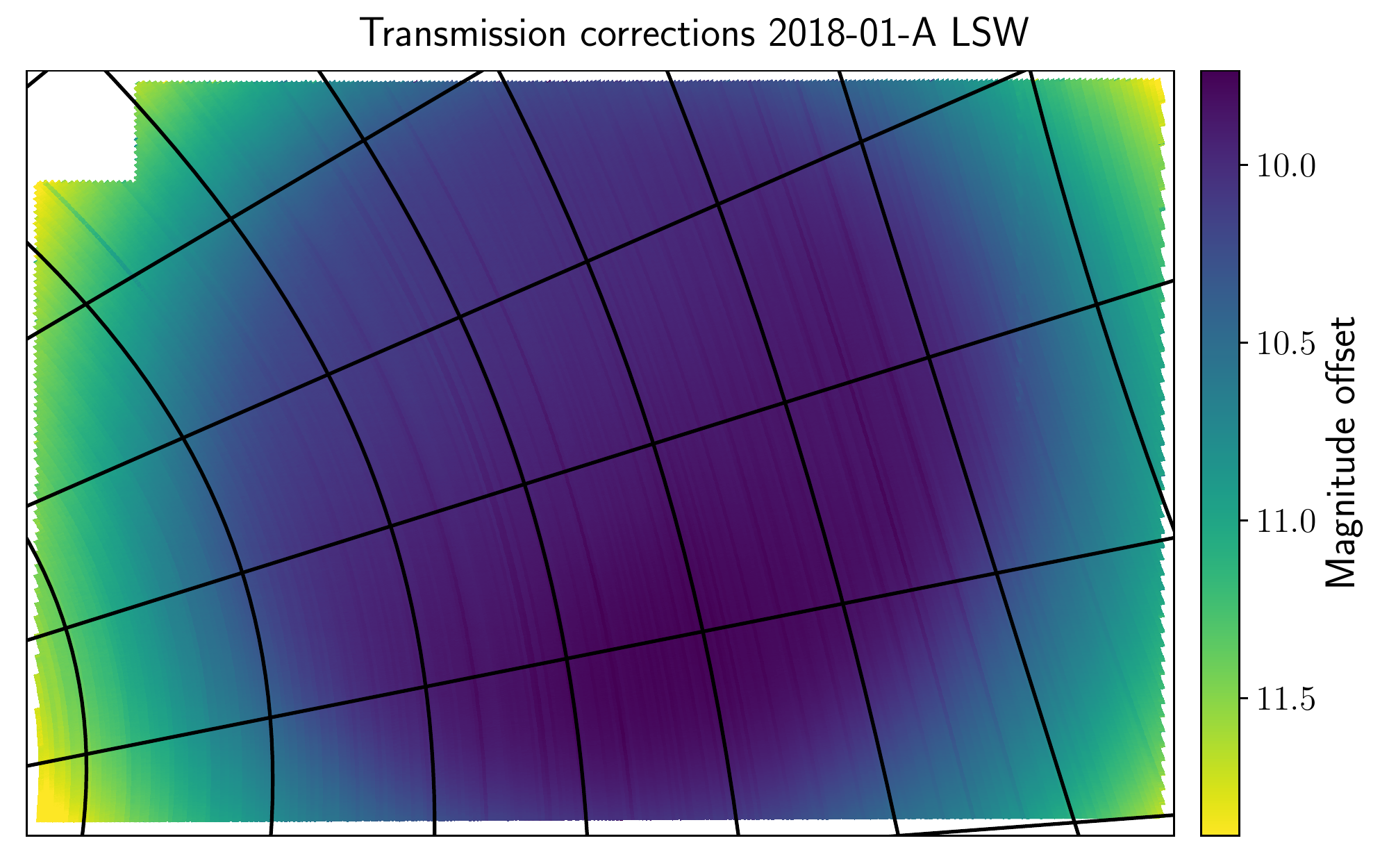}
  \includegraphics[width=8.5cm]{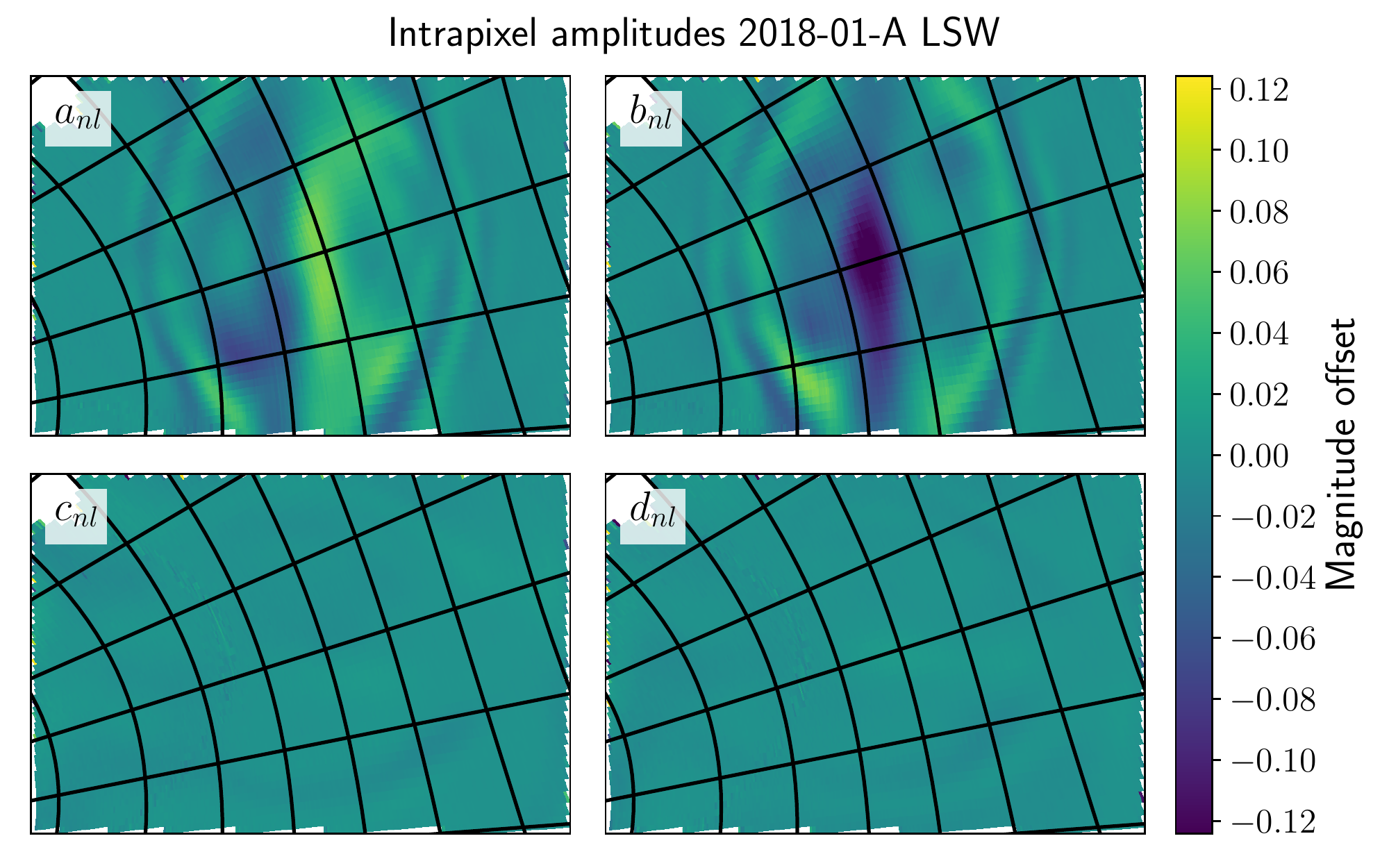}
  \caption{The transmission and intrapixel amplitude maps obtained from running the primary calibration on the 2018-01-A baseline of the LSW camera. The maps are shown projected onto the CCD and a grid in hour angle and declination is overlaid on top.}
  \label{fig:spatial_maps_lsw}
\end{figure*}

\begin{figure*}
  \centering
  \includegraphics[width=8.5cm]{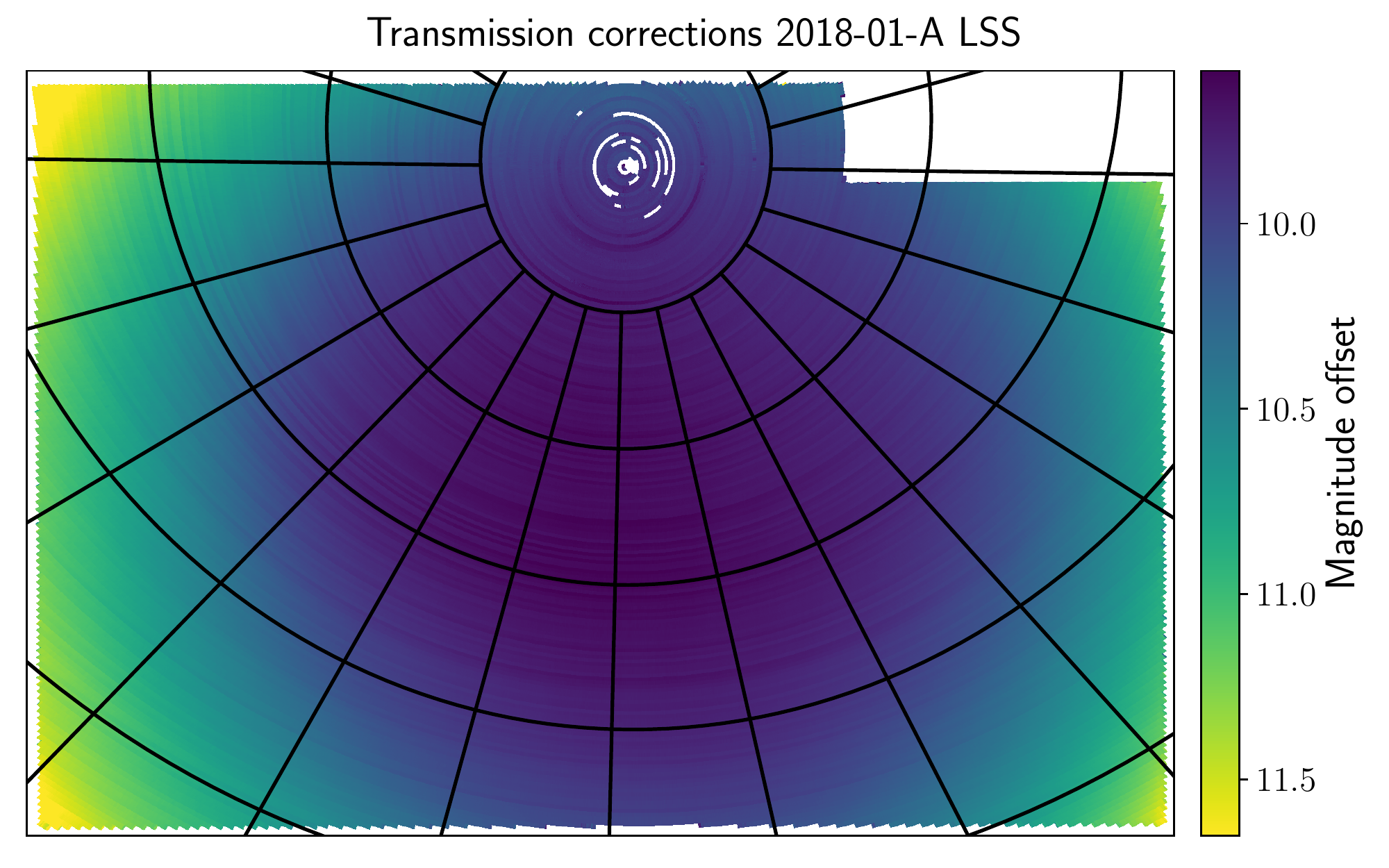}
  \includegraphics[width=8.5cm]{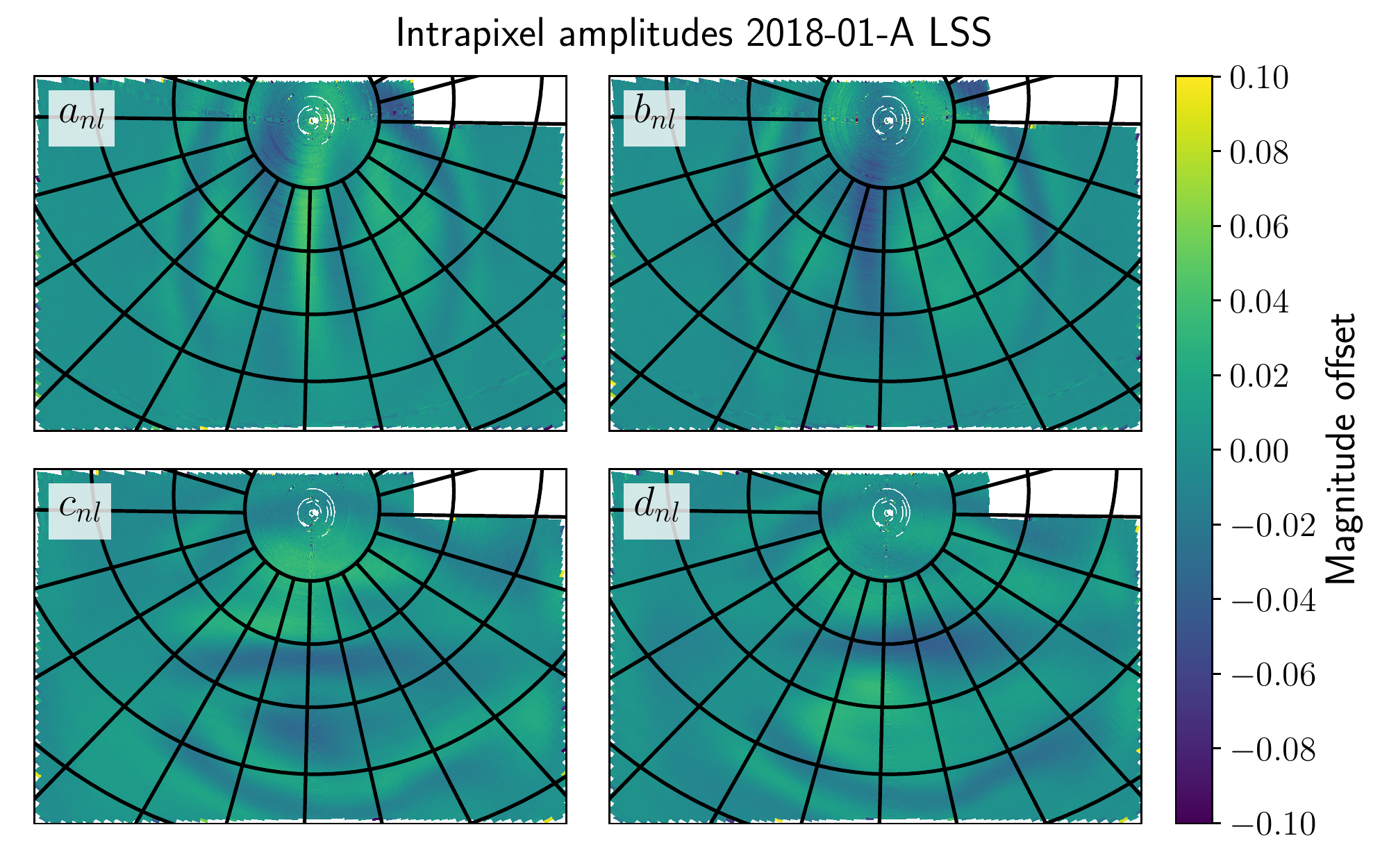}
  \caption{Same as Fig. \ref{fig:spatial_maps_lsw} for the 2018-01-A baseline of the LSS camera.}
  \label{fig:spatial_maps_lss}
\end{figure*}

\begin{figure*}
  \centering
  \includegraphics[width=8.5cm]{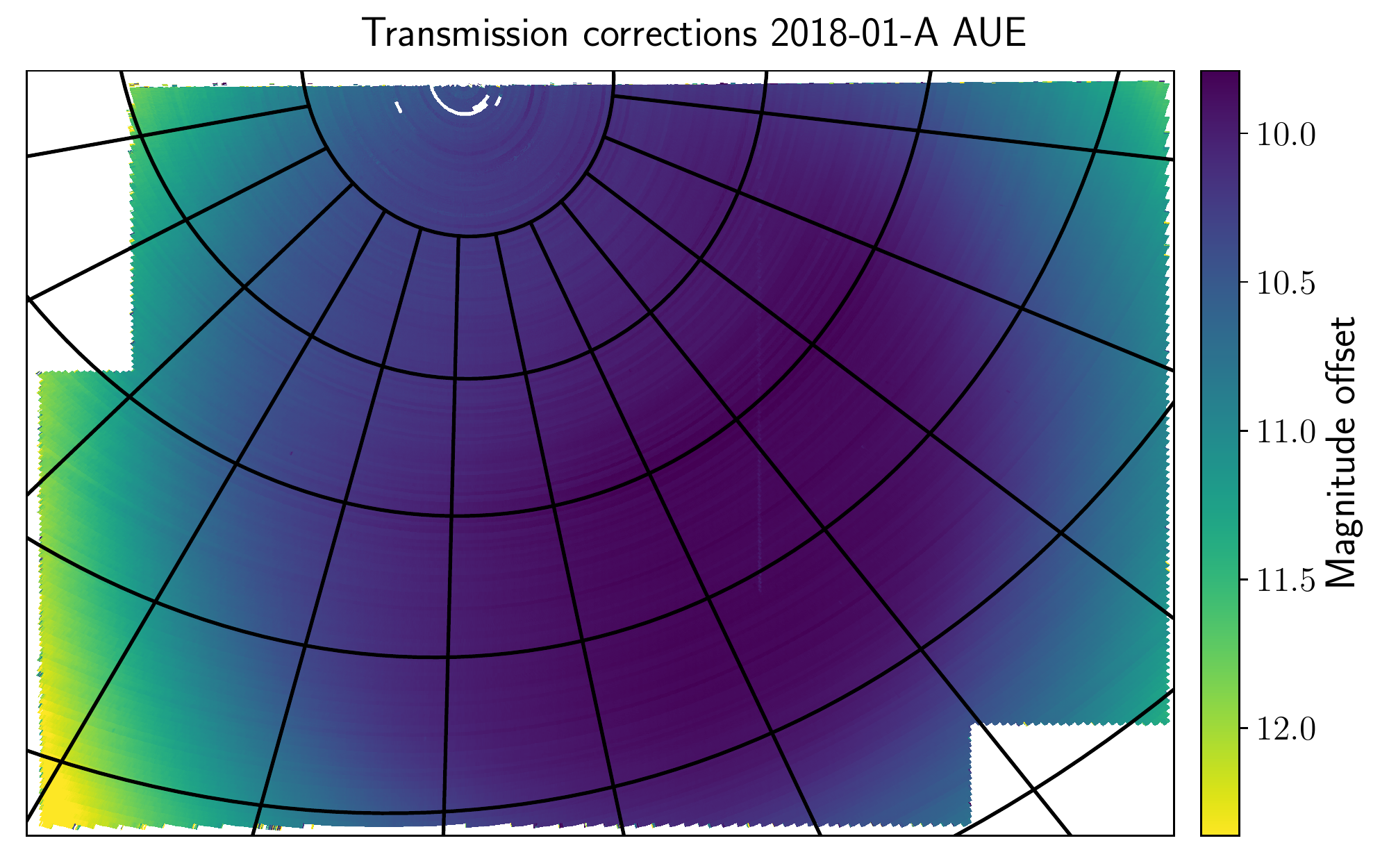}
  \includegraphics[width=8.5cm]{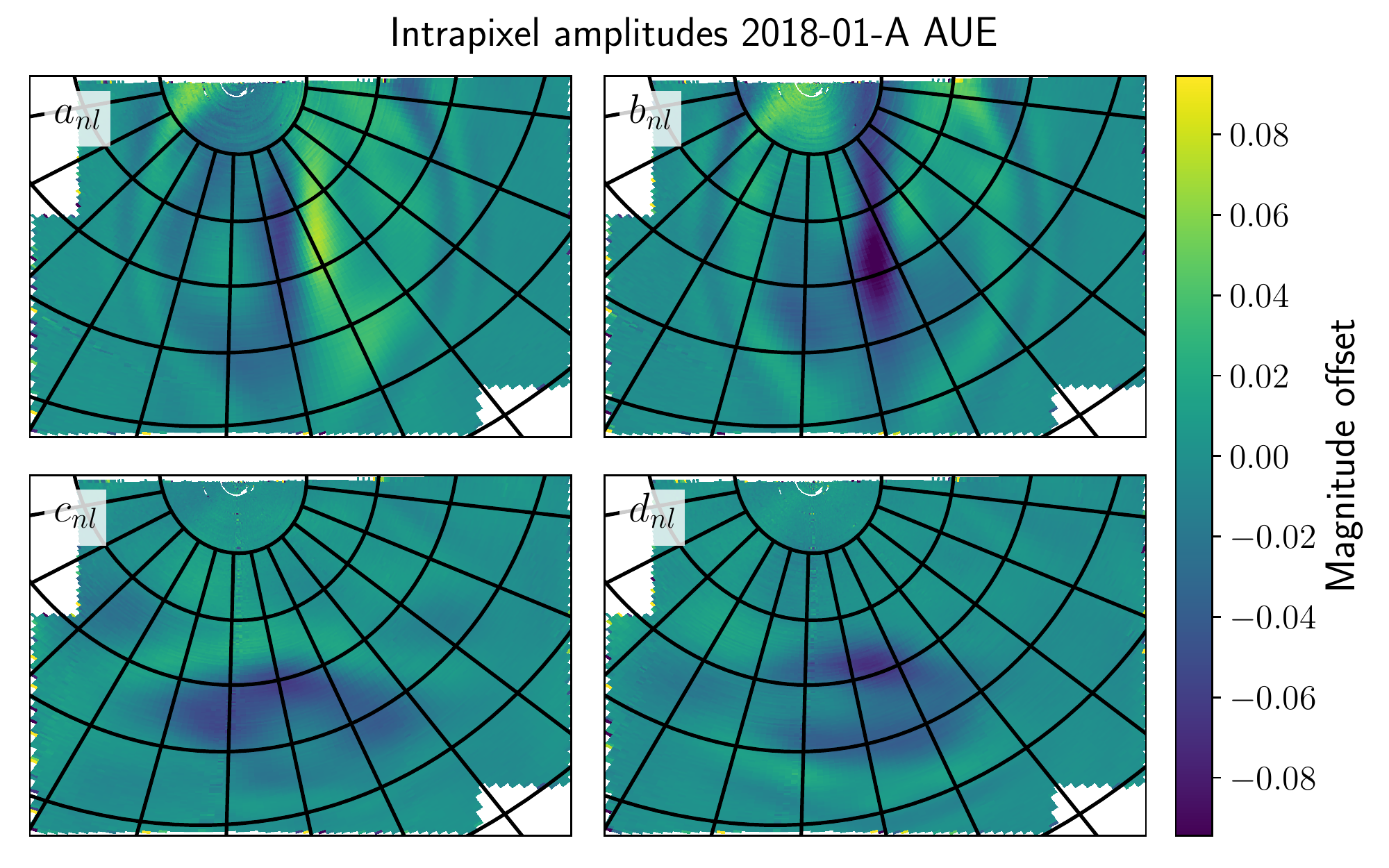}
  \caption{Same as Fig. \ref{fig:spatial_maps_lsw} for the 2018-01-A baseline of the AUE camera.}
  \label{fig:spatial_maps_aue}
\end{figure*}

Figures \ref{fig:spatial_maps_lsw}, \ref{fig:spatial_maps_lss} and \ref{fig:spatial_maps_aue} show the transmission and intrapixel amplitude maps for the LSW, LSS and AUE cameras, respectively. We selected these cameras in order to show a representative range of on-sky orientations (together with the maps shown for the LPC camera in Sect. \ref{sec:primary_cal}), and show the effect of local obscurations. The on-sky orientation of the LSW camera is representative of the LPE, LPW, LSE and LSW cameras of the MASCARA stations. For these cameras stars move approximately diagonally across the short axis of the CCDs. For this particular camera the intrapixel variations depend primarily on the $x$ position, with almost no dependence on the $y$ position, and poor sampling of $x$ and $y$ results in a decrease in quality near the bottom left of panels (a) and (b) and an arc on the left side of panels (c) and (d). The top-left corner of the CCD is masked because of the presence of a radio mast in that location. The on-sky orientation of the LSS camera is representative of the LPN and LSS cameras of the MASCARA stations. These cameras contain the poles and stars move across the CCDs in arcs. For this particular camera the intrapixel variations depend equally on the $x$ and $y$ positions and the quality of the solution decreases in an arc at the bottom of panels (a) and (b) and a vertical line down the centre of panels (c) and (d). The top-right corner of the CCD is masked to remove the MarLy telescope. The on-sky orientation of the AUE camera is representative of all bRing cameras and is qualitatively similar to LPN and LSS, though it sees less of the pole. For this particular camera the intrapixel variations depend more strongly on the $x$ position, though a clear dependence on the $y$ position is also present, and the quality of the solution decreases in an arc at the bottom of panels (a) and (b) and a diagonal down the centre of panels (c) and (d). The top-left and bottom-right corners are masked to remove obscurations by a building and a tree.

\end{document}